\documentclass[12pt,english,aps,preprint,showpacs,nofootinbib,letterpaper,fleqn]{revtex4}
\usepackage[T1]{fontenc}
\usepackage[latin1]{inputenc}
\usepackage{geometry}
\usepackage{array}
\usepackage{color}
\usepackage{graphicx}

\makeatletter

\providecommand{\tabularnewline}{\\}

\usepackage{array}
\usepackage{amsmath}
\usepackage{graphicx}
\usepackage{amssymb}
\usepackage{babel}
\makeatother

\allowdisplaybreaks[1]

\usepackage{babel}
\makeatother
\begin{document}
\newcommand{\BKM}[2]{\left<#1\!-\!|#2+\right>}

 \newcommand{\BKP}[2]{\left<#1\!+\!|#2-\right>}

 \newcommand{\BSKM}[3]{\left<\!#1\!-\!|\not\!{\!#2}_{+}|#3-\right>}

 \newcommand{\BSKP}[3]{\left<#1\!\!+\!|\!\not{\!#2}_{-}|#3+\right>}

 \newcommand{\BSSKM}[4]{\left<#1\!-\!|\not{\!#2}_{+}\not{\!#3}_{-}|#4+\right>}

 \newcommand{\BSSKP}[4]{\left<#1\!\!+\!|\not{\!#2}_{-}\not{\!#3}_{+}|#4-\right>}

 \newcommand{\g}{g_{h\gamma\gamma}}

 \newcommand{\gtilde}{\tilde{g}_{h\gamma\gamma}}

 \newcommand{\proc}{e^-e^+\to e^-e^+h}

\title{Probing the CP nature of Higgs boson through $e^{-}e^{+}\rightarrow e^{-}e^{+}\phi$}

\author{Qing-Hong Cao}

\email{qcao@ucr.edu}

\affiliation{Physics Department, University of California at Riverside, Riverside,
CA 92521, USA}

\author{F. Larios}

\email{larios@mda.cinvestav.mx}

\affiliation{Departmento de F\'{\i}sica Aplicada, CINVESTAV-M\'{e}rida, AP 73
Cordemex, 97310 M\'{e}rida, Yucat\'{a}n, Mexico}

\author{G. Tavares-Velasco}

\email{gtv@fcfm.buap.mx}

\affiliation{Facultad de Ciencias F\'{\i}sico Matem\'{a}ticas, Benem\'{e}rita
Universidad Aut\'{o}noma de Puebla, Apartado Postal 1152, Puebla,
Pue., M\'{e}xico}

\author{C.--P. Yuan}

\email{yuan@pa.msu.edu}

\affiliation{Department of Physics $\&$ Astronomy, Michigan State University,
East Lansing, Michigan 48824, USA}

\begin{abstract}
We study the production of Higgs boson ($\phi$) via the $e^{-}e^{+}\rightarrow e^{-}e^{+}\phi$
process at a future International Linear Collider (ILC) with$\sqrt{S}=1\,$TeV.
At this energy the $ZZ$-fusion rate, as predicted by the Standard Model,
 dominates over the associated
$ZH$ production rate. Here, we
consider a class of theory models based on the effective $\phi VV$
(with $V=Z,\gamma$) couplings generated by dimension-6 operators.
We also show how to use the kinematical
distributions of the final state leptons to discriminate the CP-even
and CP-odd Higgs bosons produced at the ILC, 
when the $\phi\gamma\gamma$ coupling dominates the Higgs boson 
production rate.
\end{abstract}


\maketitle

\section{Introduction}

Currently the only missing piece of the Standard Model (SM) is the
Higgs boson, which plays a vital role in this theory as it is assumed
to be responsible for the mechanism of electroweak symmetry breaking
(EWSB). Although only a single neutral CP-even Higgs boson is present 
in the SM, more than one Higgs boson may exist in several new physics 
models. For example, the two-Higgs doublet model (THDM), 
one of the simplest extensions of the SM, predicts the existence
of three neutral and two charged Higgs bosons. In the CP-conserving
THDM, two of the neutral Higgs bosons are CP-even and the remaining
one is CP-odd. If CP violation appears in the scalar potential, the
three neutral Higgs bosons may be a mixture of the CP eigenstates. 

The hopes to observe a Higgs boson at the CERN LEP collider
faded away. The lower bound on the SM Higgs boson ($H$) mass $m_{H}\geq115$
GeV was obtained from studying the process $e^{-}e^{+}\rightarrow ZH$
after combining the data taken by the four LEP experiments.
The main Higgs boson production mechanism for a Higgs boson with mass
around a few hundred GeV at a hadron collider, such as the Fermilab
Tevatron and the CERN Large Hadron Collider (LHC), is through gluon
fusion process. For $m_{H}\leq2\, m_{W}$, the Higgs boson will decay
predominantly into a $b\bar{b}$ pair, but detecting Higgs boson via
this channel is challenging because of the large QCD backgrounds.
In this mass range the rare decay modes $H\rightarrow\gamma\gamma$
and $H\rightarrow WW^{*},\, ZZ^{*}$ are the most promising channels
for detecting a SM Higgs boson. With enough integrated luminosity,
the Tevatron Run-II is expected to offer the possibility of finding
a Higgs boson with a mass up to 130 GeV through the associated $WH$
or $ZH$ production processes. At the LHC, the gluon fusion production
mechanism will allow us to search for a Higgs boson via the $H\rightarrow\gamma\gamma$
decay mode for $m_{H}$ lower than about 130 GeV, whereas for $m_{H}$ ranging
from 130 GeV to 180 GeV the $H\rightarrow WW^{*},\, ZZ^{*}$ decay modes 
could yield an observable signal~\cite{Carena:2002es}. 

Once a Higgs boson is detected, the next task will be to find out
whether it is the SM Higgs boson or some other scalar bosons predicted
by other theories. To this aim it is crucial to determine the properties
of this particle, which may shed light on the EWSB mechanism. Therefore,
apart from a precise determination of the Higgs boson mass (or masses
if there is more than one Higgs boson), one of the main tasks of future
colliders will be to determine other properties of Higgs boson, such
as its decay rates, its couplings to other particles, its properties
of transformation under the discrete symmetries C (charge conjugation)
and P (parity). Several methods have been proposed in the literature 
to determine the CP properties of a Higgs boson at hadron~\cite{Gonzalez-Garcia:1998wn,Kauffman:1998yg,Plehn:2001nj,Field:2002gt,Choi:2002jk,Figy:2004pt},
lepton~\cite{Barger:1993wt,Kilian:1996wu,Minkowski:1997cv,Hagiwara:2000tk,Grzadkowski:2000hm,Grzadkowski:2000xs,Gunion:2003fd,Barger:2003rs,Desch:2003rw,Dova:2004cj,Rouge:2005iy,Ellis:2005ik,Biswal:2005fh,Han:2000mi}
and photon~\cite{Grzadkowski:1992sa,Gounaris:1997ef,Ellis:2004hw,Niezurawski:2004ga}
colliders. Some of these methods rely on the study of certain energy
and angular correlations of the decay products of the Higgs boson,
whereas others concentrate on the change in its production rate due
to the structure of its couplings to gauge bosons or heavy fermions. 
For instance, it
was shown that polarized photon beams could be used to determine the
CP property of a Higgs boson produced at resonance via photon
fusion process~\cite{Grzadkowski:1992sa}. Also, the spin and parity
properties of the Higgs boson were analyzed for the $e^{-}e^{+}\rightarrow ZH$
process~\cite{Barger:1993wt}. In this work we show that
the $e^{-}e^{+}\rightarrow e^{-}e^{+}\phi$ process through $VV$
(with $V=Z,\gamma$) fusion at the future International Linear Collider
(ILC) could help us to determine the CP nature of the Higgs boson.
It has long been known that for large center-of-mass energies (greater than about 
500 GeV) the $ZZ$ fusion channel becomes dominant in the $e^{-}e^{+}\rightarrow e^{-}e^{+}\phi$
process~\cite{Gunion:1998jc}. 
In {Ref.~\cite{Han:2000mi}}, it was shown how to measure 
the degree of CP violation in the Higgs sector via 
$ZZ$ fusion process at a $\sqrt{S}=500\,{\rm GeV}$ ILC.
As a complementary study, we will not consider CP violating cases in this work.
We shall assume CP invariant interactions and study how to determine the 
CP property of the Higgs boson produced at the ILC 
with a higher center-of-mass energy ($\sqrt{S}=1\,{\rm TeV}$).
Since at this energy, $ZZ$ fusion contributes with $95\%$ of the total
cross section, as predicted by the SM, we choose to focus on the
$VV$ fusion channel alone. 

Rather than focusing on some specific model and calculating 
the respective loop contributions to the $\phi VV$ vertex, our study is
based on the general Lorentz invariant scalar-vector boson vertex.
In general, three possible couplings $\phi ZZ$, $\phi Z \gamma$ and
$\phi\gamma\gamma$, may give rise to $ZZ$, $Z\gamma$ and $\gamma\gamma$
fusion processes, which at high energies will become the main source
of Higgs boson production. As we shall see later, we found that the
effects of the $\phi\gamma\gamma$ couplings are more effective for
our task, as they can give large contributions to $e^{-}e^{+}\phi$
production due to collinear enhancement. 

The rest of this work is organized as follows. In Section II we present
a brief description of the general couplings. Section III is devoted
to a Monte Carlo analysis for the $e^{-}e^{+}\rightarrow e^{-}e^{+}\phi$
process. Both polarized and unpolarized beams are considered and special
attention is paid to those kinematical distributions which are sensitive
to both the strength of the anomalous $\phi VV$ coupling and the
CP property of the Higgs boson. Our conclusions will be presented
in Section IV. Finally, the helicity amplitudes for the $e^{-}e^{+}\rightarrow e^{-}e^{+}\phi$
process are presented in Appendix A for completeness.

\section{Anomalous $\phi VV$ coupling}

The general Higgs-gauge boson vertex describing $\phi\gamma\gamma$,
$\phi  Z \gamma$ and $\phi ZZ$ couplings can be written as: \begin{equation}
\Gamma_{\phi VV}^{\mu\nu}=i\, a{}_{\phi ZZ}\, g{}^{\mu\nu}\;+\; i\,\frac{g_{\phi VV}}{v}\,\left(\, p_{2}^{\mu}p_{1}^{\nu}-g^{\mu\nu}\, p_{1}\cdot p_{2}\,\right)\;+\; i\,\frac{\widetilde{g}_{\phi VV}}{v}\,\varepsilon^{\mu\nu\rho\sigma}\,{p_{1}}_{\rho}{p_{2}}_{\sigma} \, ,\label{phi-gamma-Z}\end{equation}
 where $p_{1}^{\mu}$ and $p_{2}^{\nu}$ are the momenta of the gauge
bosons. The $a_{\phi ZZ}\, g{}^{\mu\nu}$term represents the 
dimension-4 coupling that appears in models such as the SM, the THDM 
and the MSSM. For simplicity, we will use the SM value $a_{\phi ZZ}=2\, M_{Z}^{2}/v$
in our study. 

The anomalous $\phi VV$ (with $V=\gamma,V$) couplings with dimension
more than four could arise at the one-loop level in a renormalizable
theory. Also, from the standpoint of the effective Lagrangian approach
(ELA)~\cite{Weinberg:1978kz,Georgi:1991ch}, these couplings can
be induced by dimension 6 (and higher) operators~\cite{Buchmuller:1985jz,Arzt:1994gp}.
For example, an effective Lagrangian using a $SU(2)\times U(1)$ scalar
doublet that couples with the gauge bosons through dimension-6 operators
would be: \begin{equation}
\mathcal{L}_{eff}^{\phi VV}=\frac{1}{\Lambda^{2}}\sum_{i}\left(f_{i}
\mathcal{O}_{i}+\tilde{f}_{i}\tilde{\mathcal{O}_{i}}\right) \, ,\label{eq: LagphiAA}\end{equation}
where the CP-conserving operators $\mathcal{O}_{i}$ are given by~\cite{Hagiwara:1993qt}\begin{eqnarray*}
\mathcal{O}_{BW} & = & \Phi^{\dagger}{\textrm{B}}_{\mu\nu}{\textrm{W}}^{\mu\nu}\Phi,\\
\mathcal{O}_{WW} & = & \Phi^{\dagger}{\textrm{W}}_{\mu\nu}{\textrm{W}}^{\mu\nu}\Phi,\\
\mathcal{O}_{BB} & = & \Phi^{\dagger}{\textrm{B}}_{\mu\nu}{\textrm{B}}^{\mu\nu}\Phi,\end{eqnarray*}
 in which $\Phi$ is the Higgs doublet, ${\textrm{B}}_{\mu\nu}=ig^{\prime}\, B_{\mu\nu}$
and ${\textrm{W}}_{\mu\nu}=ig/2\,\sigma^{a}W_{\mu\nu}^{a}$, with $B_{\mu\nu}$
and $W_{\mu\nu}^{a}$ the strength tensors of the gauge fields;
$g'$ and $g$ are U(1) and SU(2) gauge couplings, respectively. As for
the CP-violating operators $\widetilde{\mathcal{O}}_{i}$, they are
obtained from the above ones after replacing either of the two strength
tensors by their respective dual~\cite{Gounaris:1997ef}. 
The simultaneous presence of both $g_{\phi VV}$ and $\widetilde{g}_{\phi VV}$
for one boson would signal CP violation. As mentioned, we will
not consider that case in this work, for this we refer the reader
to Ref.~\cite{Han:2000mi}. In our study we assume a SM-like CP-even
Higgs boson $h$ with anomalous couplings $g_{hVV}$ and/or a CP-odd
Higgs boson $A$ of same mass but with couplings $\tilde{g}_{AVV}$~%
\footnote{In general, the CP-odd Higgs boson can have a mass different from
the CP-even Higgs boson. Since the goal of this study is to develop 
methods for determining the CP property of the Higgs boson, we assume 
a similar mass for these Higgs bosons to simplify our comparison.}
We will also refer to the SM Higgs boson $H$ for the purpose of comparison.
Throughout this paper, we refer the production of the SM Higgs boson $H$ as 
the one generated from the Born level dimension-4 $HZZ$ coupling. Even though 
higher dimension anomalous operators may arise from loop corrections, 
we will assume their effects to be negligible. In contrast, for the production 
of the SM-like Higgs boson $h$, we include both the 
Born level dimension-4 $HZZ$ coupling and the higher dimension anomalous 
 $g_{hVV}$ couplings in our calculations. For the production of the 
CP-odd Higgs boson $A$, only the higher dimension anomalous 
 ${\tilde g}_{AVV}$ couplings are included.

A common approach for determining the CP property of Higgs boson 
is to examine the $Z\phi$ associated production
($e^{-}e^{+}\rightarrow Z\phi\to f^{-}f^{+}\phi$) and analyze the
angular distribution of the $Z$ decay products~\cite{Barger:1993wt}.
Evidently, for this approach the relevant coupling is $\phi ZZ$ (or
$\phi Z\gamma$). In this work we will consider a higher energy of
1 TeV for the $e^{+}e^{-}$ collider and examine the 
$e^{-}e^{+}\rightarrow e^{-}e^{+}\phi$ process.
 At this energy $VV$ fusion
($V=Z,\gamma$) becomes dominant. Eventually, we will focus on the
$\gamma\gamma$ fusion sub-process which gives a substantial production
rate and could be analyzed to study the CP nature of the Higgs boson $\phi$.

In order to make any predictions for those processes involving the
effective $\phi VV$ coupling, it is necessary to know the order of
magnitude of the effective coupling constants $g_{\phi VV}$. Several
processes have been discussed in the literature that can be used to
obtain bounds on these coefficients. In particular, based on the ELA,
the constraints from the LEP and the Tevatron on the reactions $e^{+}e^{-}\rightarrow\gamma\gamma\gamma$,
$p\bar{p}\rightarrow jj\gamma\gamma$, $p\bar{p}\rightarrow\gamma\gamma+E_{T}$,
and $p\bar{p}\rightarrow\gamma\gamma\gamma$ are known to yield bounds
on $f_{i}/\Lambda^{2}$~\cite{Hagiwara:1993qt,Gonzalez-Garcia:1998wn}.
To be consistent with the bounds reported in Ref.~\cite{Gonzalez-Garcia:1998wn}
we will consider $|f_{i}|/\Lambda^{2}\leq50\;{\textrm{TeV}}^{-2}$,
which implies $\left|g_{\phi VV}\right|\leq0.3$ in our notation,
cf. Eq.~(\ref{phi-gamma-Z}).

\section{Monte Carlo analysis for $e^{-}e^{+}\rightarrow e^{-}e^{+}\phi$}

We now turn to discuss how the $e^{-}e^{+}\rightarrow e^{-}e^{+}\phi$
process can be used to determine the CP property of the Higgs boson.
In the following, $H$ stands for the SM Higgs boson, and as discussed 
above, $h$ and $A$ denote the CP-even and the CP-odd Higgs boson,
respectively. 

\begin{figure}
\includegraphics[scale=0.5]{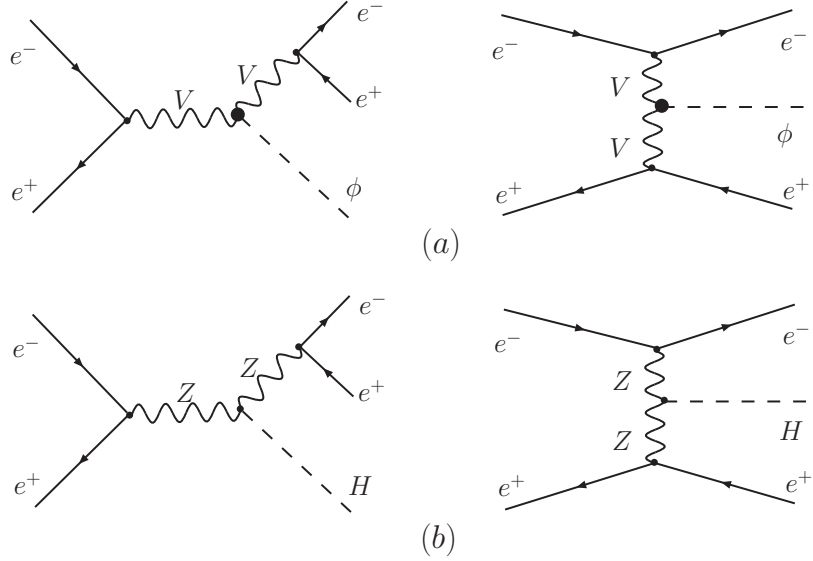}

\caption{Feynman diagrams for $e^{-}e^{+}\rightarrow e^{-}e^{+}\phi$: (a)
the effective theory where the heavy dot denotes the effective $\phi VV$
coupling ($V=\gamma,Z$); (b) the Standard Model. \label{feynman}}
\end{figure}

The Feynman diagrams for $e^{-}e^{+}\rightarrow e^{-}e^{+}\phi$ are
shown in Fig.~\ref{feynman}. For the SM Higgs boson $H$, the contributions
arise from Higgstrahlung (the left diagram) and vector-boson fusion
(the right diagram) processes. For the CP-even Higgs boson $h$, in
addition to the SM diagrams there would be contributions arising from
the higher dimension $hVV$ couplings. In contrast, for the CP-odd
Higgs boson $A$, the contributions arise only from the higher 
dimension $AVV$ operators,
as there is no SM-like tree level $AZZ$ coupling. 

We will carry out a Monte Carlo study and construct some CP-odd observables
which are sensitive to the strength of the anomalous $\phi VV$ couplings 
and can be used 
to discriminate a CP-even from a CP-odd Higgs
boson. In our analysis we will consider both unpolarized and polarized
lepton beams, and the convention for the momenta of the external particles
is taken as follows \begin{equation}
e^{-}(p_{1})e^{+}(p_{2})\rightarrow e^{-}(p_{3})e^{+}(p_{4})\phi(p_{5}) \, ,\label{momenta}\end{equation}
where $p_1$ and $p_2$ are the incoming momenta and $p_3$, $p_4$ and $p_5$ are 
out-going momenta.

Given the effective interaction of Eq.~(\ref{phi-gamma-Z}),
it is straightforward to calculate the
helicity amplitudes for $e^{-}e^{+}\rightarrow e^{-}e^{+}\phi$. To
this end we used the spinor product formalism introduced 
in Refs.~\cite{Hagiwara:1985yu,Carlson:1995ei}.
The resulting amplitudes are presented in Appendix A for the sake
of completeness. To numerically illustrate the method of our analysis, 
we will take the Higgs boson mass to be 115
GeV (for each $\phi\,=\, H,h,A$) and the center-of-mass energy of the
$e^{+}e^{-}$ collider to be 1 TeV. 

We will discuss first the case of unpolarized beams.

\subsection{Initial considerations and basic cuts for the $\gamma\gamma$ fusion
$t$-channel diagram}

The $e^{-}e^{+}\rightarrow e^{-}e^{+}\phi$ processes can be divided
into two sub-processes: the Higgstrahlung process ($V\phi$ production
with subsequent decay of the vector boson $V$ into the pair of charged
leptons) and the vector-boson fusion process. The Higgstrahlung process
is $s$-channel like while the vector-boson fusion process is $t$-channel
like. We can separate them by examining the invariant mass of the
outgoing $e^{-}e^{+}$ pair. (Of course, for $Z\phi$ production one
can include the other decay modes of the $Z$ boson which are not
considered in this study). In the Higgstrahlung process, the two 
final state charged
leptons originate from the vector-boson decay. Therefore, if we require
the invariant mass ($M_{34}=m_{e^{+}e^{-}}$) to be within a $10\,{\rm {\rm GeV}}$
range of the $Z$-boson mass, the cross section will come mostly from
the $s$-channel diagram. Otherwise, the main contribution will come
from the $t$-channel ($ZZ$ fusion) diagram.

\begin{table}

\caption{The unpolarized cross sections of $e^{-}e^{+}\rightarrow e^{-}e^{+}\phi$,
where $\phi$ is either the SM Higgs boson $H$, the CP-even boson
$h$ or the CP-odd $A$. We have taken $\sqrt{S}=1\,{\rm TeV}$, $m_{\phi}=115\,{\rm GeV}$,
$g_{hVV}=0.1$ and $\tilde{g}_{AVV}=0.1$ for $V=Z$ or $\gamma$.
(Only one anomalous coupling is taken to be non-zero for each case.) 
We have imposed the basic cuts of Eq.~(\ref{cuts1}) in order to avoid
the divergent collinear behavior of the scattering processes arising from 
the $\phi\gamma\gamma$
couplings. \label{tableallcouplings}}

\begin{tabular}{c|c|c}
\hline 
process&
$\sigma$(fb)&
$\sigma$(fb)\tabularnewline
&
$\left|M_{34}-m_{Z}\right|<10\,{\rm GeV}$&
$\left|M_{34}-m_{Z}\right|>10\,{\rm GeV}$ \tabularnewline
\hline 
$e^{-}e^{+}\rightarrow e^{-}e^{+}H$&
0.42&
9.0\tabularnewline
\hline
$e^{-}e^{+}\rightarrow e^{-}e^{+}h$ ($g_{hZZ}$)&
 0.88 &
 8.6 \tabularnewline
\hline
$e^{-}e^{+}\rightarrow e^{-}e^{+}h$ ($g$$_{h\gamma\gamma}$)&
0.44&
13.9\tabularnewline
\hline
$e^{-}e^{+}\rightarrow e^{-}e^{+}A$ ($\tilde{g}$$_{AZZ}$)&
 0.21 &
 0.17  \tabularnewline
\hline
$e^{-}e^{+}\rightarrow e^{-}e^{+}A$ ($\tilde{g}$$_{A\gamma\gamma}$)&
0.02&
4.22\tabularnewline
\hline
\end{tabular}
\end{table}

On the other hand, both $s$- and $t$-channel diagrams may have a
singular behavior when the $\phi\gamma\gamma$ (or $\phi  Z \gamma$)
coupling is considered. In particular, collinear and 
nearly on-shell photon
divergences appear, and we must impose appropriate cuts 
to avoid these kinematical regions. In this work we will start by imposing the following
conditions: \begin{eqnarray}
|\cos\theta_{3}|,|\cos\theta_{4}| &  & \leq0.99\,,\nonumber \\
P_{T3},P_{T4} &  & \geq10\;{\textrm{GeV}}\,,\label{cuts1}\\
M_{34} &  & \geq5\;{\textrm{GeV}}\,.\nonumber \end{eqnarray}
 Here, the angle $\theta_{3}$ ($\theta_{4}$) is the polar angle
of the outgoing electron (positron). The incoming electron is moving
in the $+\hat{z}$ axis direction. $P_{t}$ denotes the transverse
momentum. The first two cuts are required to avoid collinear divergences
in the $t$-channel process, whereas the third cut is required to avoid the
divergence when the $s$-channel photon goes on-shell when ignoring
the electron mass in the helicity amplitude calculations. Furthermore,
the requirement on the cosine of the polar angles $|\cos\theta|\leq0.99$,
which corresponds to $172^{\circ} \geq \theta \geq 8^{\circ}$, will ensure that the
final state leptons fall within the detector coverage at the ILC.

The cross section for $e^{-}e^{+}\rightarrow e^{-}e^{+}\phi$ with
unpolarized beams and a fixed value of the coupling coefficients is
shown in Table~\ref{tableallcouplings}. The SM prediction for the
$e^{-}e^{+}\rightarrow ZH\rightarrow e^{-}e^{+}H$ process is about
0.4 fb, much smaller than the 9 fb predicted for the $t$-channel
$ZZ$ fusion process at $\sqrt{S}=1\,{\rm TeV}$. The purpose of this
work is to distinguish between the CP-even and CP-odd scalars, and
we will concentrate on studying the $t$-channel process. Table~\ref{tableallcouplings}
also shows that the contribution from the $\phi\gamma\gamma$ coupling
is potentially larger than the contribution from the anomalous $\phi ZZ$
coupling with the same magnitude. This is because the collinear
divergences mentioned above only appear in scattering processes with nearly 
on-shell photon.

Apart from analyzing the usual kinematic distributions, we will also
consider the following angular variables which will be shown later to be
sensitive to the CP nature of Higgs boson: 
\begin{eqnarray}
X_{e^{-}e^{+}} & \equiv & X\;\equiv\;(\hat{p}_{3}-\hat{p}_{4})\cdot\hat{p}_{1}\;=\cos\theta_{3}-\cos\theta_{4},\nonumber \\
Y_{e^{-}e^{+}} & \equiv & Y\;\equiv\;\frac{\hat{p}_{3}\cdot(\vec{p}_{1}\times\vec{p}_{4})}{|\vec{p}_{1}\times\vec{p}_{4}|}\;-\;\frac{\hat{p}_{4}\cdot(\vec{p}_{1}\times\vec{p}_{3})}{|\vec{p}_{1}\times\vec{p}_{3}|},\label{eq:xydefintion}\end{eqnarray}
 where the unit vectors $\hat{p}_{i}=\vec{p}_{i}/|\vec{p}_{i}|$. Generally, the
experimental analysis of these variables would require particle identification
of the outgoing leptons. However, the absolute values of
$X_{e^{-}e^{+}}$ and $Y_{e^{-}e^{+}}$ could also be used with the
advantage of not requiring to determine the charge of the outgoing
leptons, separating $e^{+}$ from $e^{-}$ in the final state, to
discriminate a CP-even from a CP-odd Higgs boson.

\subsection{Comparison of production rates from the $\phi\gamma\gamma$, $\phi  Z \gamma$
and $\phi ZZ$ couplings in the $t$-channel process}

For the rest of this work we focus our attention on the $t$-channel
process, so that we will require the $e^{-}e^{+}$ invariant mass
to be away from the $Z$ mass value: \begin{eqnarray}
\left|M_{34}-m_{Z}\right|\geq10\;{\textrm{GeV}} \, . \label{cutzpeak}\end{eqnarray}
 It is understood that the conditions of Eqs.~(\ref{cuts1})~and~(\ref{cutzpeak})
are being applied to the following analysis. We will refer to these
conditions as \textit{basic cuts}. 

\begin{figure}
\includegraphics[clip,scale=0.5]{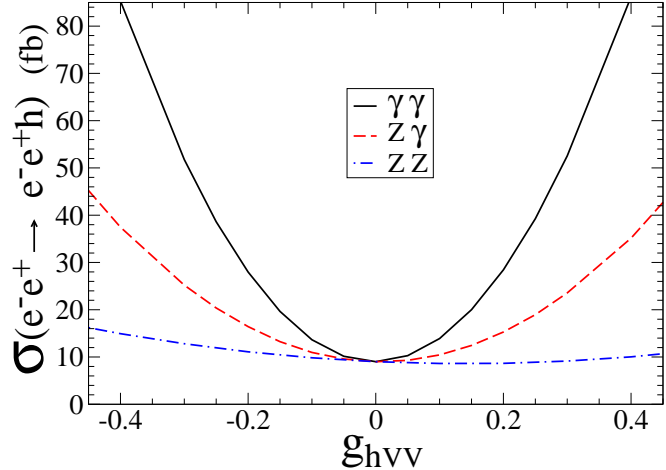}

\caption{The unpolarized $e^{-}e^{+}\rightarrow e^{-}e^{+}h$ production cross section
for different values of the 3 CP-even coupling coefficients: $g_{h \gamma\gamma }$,
$g_{h  Z  \gamma }$ and $g_{hZZ}$. The \textit{basic
cuts}, Eqs.~(\ref{cuts1}) and (\ref{cutzpeak}), have been applied, and 
only one anomalous coupling is taken to be non-zero for each case.
\label{crossec} }
\end{figure}

In Fig.~\ref{crossec} we show the $e^{-}e^{+}\rightarrow e^{-}e^{+}h$
production cross section induced by each of the 3 couplings, $g_{h \gamma\gamma}$,
$g_{h  Z  \gamma }$ and $g_{hZZ}$. Only one coupling is taken to be non-zero
for each case. Because of the collinear divergent behavior originated
from the photon propagator, the contribution from the $h\gamma\gamma$
coupling is much larger for similar values of the coupling coefficients. 

For the case of $e^{-}e^{+}\rightarrow e^{-}e^{+}A$ production, the
cross section depends directly on the $\tilde{g}_{AVV}$ coupling.
We can write down the numerical value of the production cross section
as: \begin{eqnarray}
\sigma(A \gamma\gamma) & = & 421\times\tilde{g}_{A \gamma\gamma}^{2}\;\;({\textrm{fb}}) \, ,\nonumber \\
\sigma(A Z \gamma) & = & 145\times\tilde{g}_{A \gamma Z}^{2}\;\;({\textrm{fb}}) \, , \label{oddsigma}\\
\sigma(AZZ) & = & 21.4\times\tilde{g}_{A ZZ}^{2}\;\;({\textrm{fb}}) \, . \nonumber \end{eqnarray}

In general, it is not possible to predict the relative sizes of the
anomalous coupling coefficients
without knowing the underlying theory. In case that the 
 underlying physics induce all three effective couplings with
similar size, the $\phi\gamma\gamma$ coupling will produce the biggest
effect on the $t$-channel production rate of Higgs boson. Thus, hereafter,
we will assume that the effect of this coupling indeed dominates the
Higgs boson production rate and we shall focus our attention on the
effect of this coupling to various kinematic distributions of the
CP-even and CP-odd Higgs boson signal events.

\subsection{Unpolarized beams}

\begin{figure}
\includegraphics[scale=0.7]{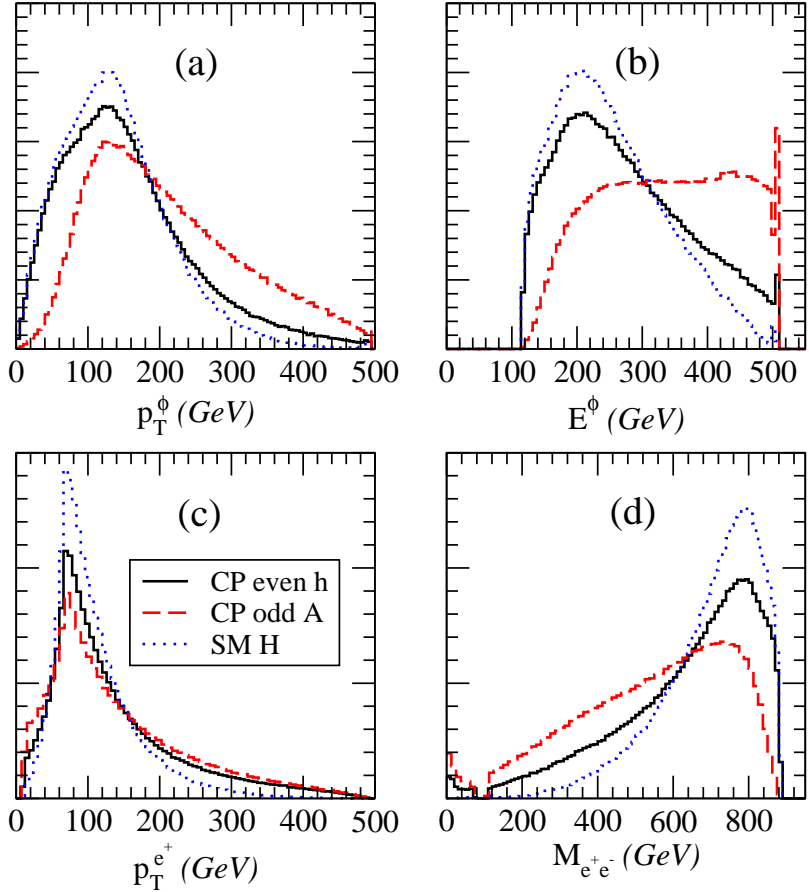}

\caption{The normalized differential cross sections for the $e^{-}e^{+}\rightarrow e^{-}e^{+}\phi$
processes as a function of the corresponding kinematical variable.
We show the distributions for the SM $H$ (blue dotted line), the
CP-even $h$ (black solid line) and the CP-odd $A$ (red dashed line)
Higgs bosons. The values $g_{h\gamma\gamma}=\tilde{g}_{A\gamma\gamma}=0.1$
have been used. The \textit{basic cuts} ( Eqs.~(\ref{cuts1})~and~(\ref{cutzpeak})
) have been applied.\label{distcuts1} }
\end{figure}

After imposing the \textit{basic cuts}, cf. Eqs.~(\ref{cuts1})~and~(\ref{cutzpeak}),
we show in Fig.~\ref{distcuts1} a few kinematic distributions which
are useful for determining the CP property of the Higgs boson. They
are the distributions of the transverse momentum and energy of the
Higgs boson and lepton, and the invariant mass of the two final state
leptons.

\begin{figure}
\includegraphics[scale=0.7]{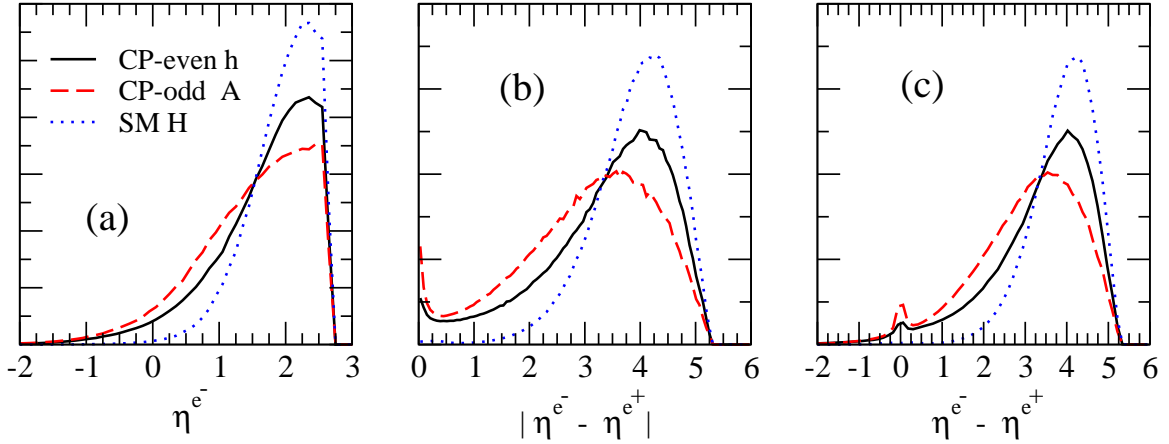}

\caption{The normalized differential cross section for the $e^{-}e^{+}\rightarrow e^{-}e^{+}\phi$
process as a function of the rapidity (difference). We show the distributions
for the SM $H$ (blue dotted line), the CP-even $h$ (black solid
line) and the CP-odd $A$ (red dashed line) Higgs bosons. The values
$g_{h\gamma\gamma}=\tilde{g}_{A\gamma\gamma}=0.1$ have been used, and 
the \textit{basic cuts} have been applied. \label{rapidiference} }
\end{figure}

The results shown in Fig.~\ref{distcuts1} can be understood as follows.
The small tail in the low mass region of the lepton pair invariant
mass plot suggests that the production rate is dominated by the $ZZ$
and/or $\gamma\gamma$ fusion processes, and the $s$-channel process
is not as important as the $t$-channel process in high energy collision
with center-of-mass energy around 1 TeV. The SM Higgs boson production
is through the dimension-4 $HZZ$ interaction only. While the CP-even Higgs boson
$h$ could be produced through either the dimension-4 $hZZ$ or 
the anomalous $h \gamma\gamma $ 
interaction, the CP-odd Higgs boson could only be produced through
the $A \gamma\gamma $ interaction. Because of the propagator of the
$Z$-boson in the $ZZ$ fusion process, the typical transverse momentum
of the final state lepton after emitting the $Z$ boson in the $ZZ$
fusion process is about half of the $Z$ boson mass. In case of the
production of the CP-odd Higgs boson via $\gamma\gamma$ fusion, the
final state lepton typically has a smaller transverse momentum,
cf. Fig.~\ref{distcuts1}(c), due to the collinear enhancement in
the emission of an effective photon from the lepton.
Since the effective photon is transversely polarized, the lepton after
emitting the effective photon in the $\gamma\gamma$ fusion process
can have a large angle relative to the beam axis, and yields a smaller
magnitude in the absolute value of its rapidity. 
In contrast, after emitting a longitudinal
(virtual) $Z$ boson, the lepton prefers to move along the forward
direction which is obvious from the consideration of helicity conservation.
The net effect of the collinear enhancement and the transverse polarization
of the emitted effective photon to the lepton rapidity distribution 
can be seen in Fig.~\ref{rapidiference}(a), which shows the rapidity
distribution of the final state electron with the choice of the incoming
electron in the $+\hat{z}$ direction. The distribution of the final
state positron is not shown because it has the same shape as the electron's
but in the negative rapidity direction. As compared to the lepton
rapidity distribution predicted by the SM, the one predicted for the
CP-odd Higgs boson shows a flatter distribution. The result for the
CP-even non-SM Higgs boson $h$ lies between the above two curves.
In Fig.~\ref{rapidiference}(b), we also show the distribution of
the rapidity difference, in its absolute value, between the two final
state leptons in the Higgs boson events. By taking only the absolute 
value of the rapidity difference between the two charged leptons,
we eliminate the requirement of identifying the electrical charge
of the two outgoing leptons, i.e., separating $e^{+}$
from $e^{-}$, which could be challenging for leptons with large rapidity
(in magnitude). For completeness, we also show in Fig.~\ref{rapidiference}(c)
the distribution of the lepton rapidity difference without taking
its absolute values (for the case that the electrical charge of the
lepton is identified). Due to the collinear enhancement discussed
above, very little event rate falls into the negative region. 

Next, let us discuss the kinematic distributions of the Higgs bosons.
As shown in Figs.~\ref{distcuts1}(a) and (b), the energy of the
CP-odd Higgs boson is typically higher than that of the SM Higgs boson
because the invariant mass of the final state lepton pairs in the
CP-odd Higgs signal events is lower than that in the SM Higgs signal
events, cf. Fig.~\ref{distcuts1}(d), which is due to the lower transverse
momentum that the final state leptons can have in the CP-odd Higgs
events. Furthermore, due to the transversity of the emitted
effective photon in the $\gamma\gamma$ fusion process, the possibility
for producing two energetic leptons from $\gamma\gamma$ fusion process
is small, therefore the energy distribution of the CP-odd Higgs boson drops
very fast in the small energy region, cf. Fig.~\ref{distcuts1}(b).
The same reasoning also explains the transverse momentum distribution
of the Higgs boson in Fig.~\ref{distcuts1}(a). In all cases, the
maximal allowed value for 
the energy of the Higgs boson 
$E^{\phi}$ is constrained by the center-of-mass
energy ($\sqrt{S}$) of the collider 
and the masses of final state particles, it is\begin{equation}
E_{{\textrm{max}}}^{\phi}=\frac{\sqrt{S}}{2}-\frac{\left(m_{e^{+}}+m_{e^{-}}\right)^{2}-m_{\phi}^{2}}{2\sqrt{S}}\sim507\,{\textrm{GeV}},\label{eq:eh_max}\end{equation}
 as can be deduced from the three-body phase space.

\begin{table}

\caption{The unpolarized cross sections of $e^{-}e^{+}\rightarrow e^{-}e^{+}\phi$,
where $\phi$ is either the SM Higgs boson $H$, the CP-even boson $h$
or the CP-odd $A$. We have taken $m_{\phi}=115\,{\rm GeV}$ and considered 
the following cases: $g_{h\gamma\gamma}=\pm0.1$
and $\tilde{g}_{A\gamma\gamma}=0.1$. The kinematic cuts listed in
each column are applied sequentially. \label{tablecutrap}}

\begin{tabular}{c|c|c|c}
\hline 
process&
(fb)&
(fb)&
(fb)\tabularnewline
&
{\textit{basic cuts}} &
$|\eta_{e^{-}}-\eta_{e^{+}}|\leq3$ GeV &
$|\eta_{e^{-}}-\eta_{e^{+}}|\leq2$ GeV \tabularnewline
\hline
\hline 
$e^{-}e^{+}\rightarrow e^{-}e^{+}H$&
9.0&
1.09&
0.17\tabularnewline
\hline
$e^{-}e^{+}\rightarrow e^{-}e^{+}h$ \, ($g_{h\gamma\gamma}=0.1$)&
 13.9 &
 3.50 &
 1.42 \tabularnewline
\hline
$e^{-}e^{+}\rightarrow e^{-}e^{+}h$ \, ($g_{h\gamma\gamma}=-0.1$)&
13.7&
3.30&
1.28\tabularnewline
\hline
$e^{-}e^{+}\rightarrow e^{-}e^{+}A$ \, ($\tilde{g}_{A\gamma\gamma}=0.1$)&
 4.22 &
 1.91 &
 0.91  \tabularnewline
\hline
\end{tabular}
\end{table}

We note that in the above kinematic distributions, the CP-odd Higgs
boson differs from the CP-even Higgs boson, but the CP-even Higgs
boson has similar distributions as the SM Higgs boson. To separate
the CP-even Higgs boson from the SM Higgs boson background, one could
make use of the rapidity distributions of the final state leptons,
as shown in Fig.~\ref{rapidiference}. Since the CP-even
Higgs boson $h$ can be produced either from $ZZ$ or $\gamma\gamma$
fusion process, the electron rapidity distribution lies in between
the SM Higgs boson and CP-odd Higgs boson. Based on the different shapes in the
rapidity distributions of $H$, $h$ and $A$, cf. Fig.~\ref{rapidiference},
we can impose further cuts to reduce the SM rate (which is referred below as 
the ``background rate'').
In Table~\ref{tablecutrap} we show the cross sections after demanding
the rapidity difference between the two leptons to be smaller than
3 and $2$, separately. The former cut, $|\eta_{e^{-}}-\eta_{e^{+}}|<3$,
increases the signal-to-background ratio by about a factor $2$ while
keeping about $25\%$ of the signal rate for the CP-even Higgs boson
for both positive and negative $g_{h\gamma\gamma}$. For the CP-odd
Higgs boson, this cut increases the signal-to-background ratio by
about a factor of $3.7$ while keeping about $45\%$ of the signal
rate. That more CP-odd Higgs boson events survive this cut is due
to the different electron rapidity distributions as discussed above.
As shown in Fig.~\ref{rapidiference}(b), it would be possible to
make a substantial reduction (by an order of magnitude) of the SM
rate by imposing a lower rapidity difference cut. We observe, nevertheless,
that for the stringent cut ($|\eta_{e^{-}}-\eta_{e^{+}}|<2$) with
couplings $g_{\phi\gamma\gamma}$ and $\widetilde{g}_{\phi\gamma\gamma}$
of order 0.1, only a small cross section of about less than $1\,{\rm fb}$
may be left. This may make it difficult to observe any events
in a more realistic analysis. We therefore choose not to use this
stronger cut. For the remainder of this work, in addition to the \textit{basic
cuts} we will apply the following cut~%
\footnote{Notice that the cuts imposed in this work do not require 
identifying the electric charge of the leptons, i.e., separating 
 $e^{+}$ from $e^{-}$.
}:\begin{equation}
|\eta_{e^{-}}-\eta_{e^{+}}|\leq3.\label{cuts3}\end{equation}
But it is worth mentioning that with a high luminosity in the upgraded
ILC, one could use a tight cut to achieve a better significance.

\begin{figure}
\includegraphics[clip,scale=0.7]{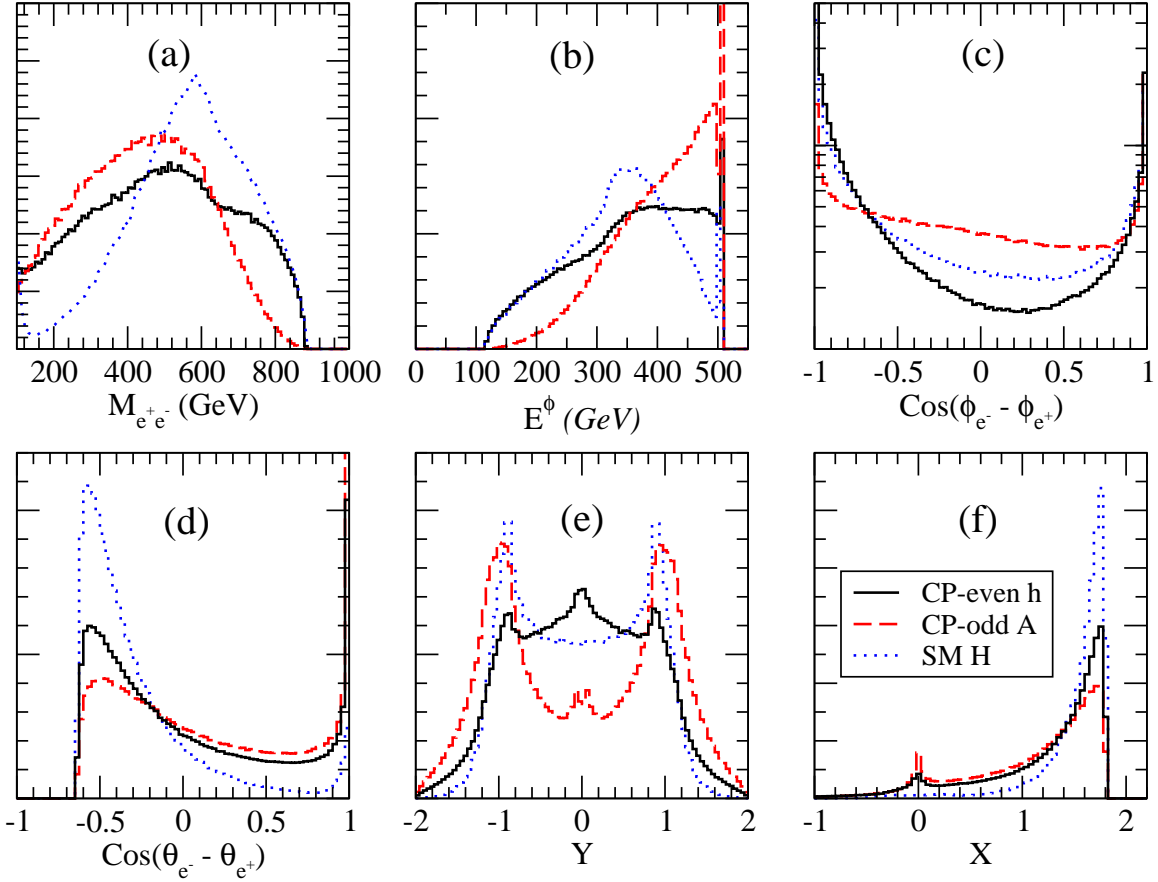}

\caption{The normalized differential cross sections for the 
unpolarized $e^{-}e^{+}\rightarrow e^{-}e^{+}\phi$
processes as a function of the corresponding kinematical variable.
We show the distributions for the SM $H$ (blue dotted line), the
CP-even $h$ (black solid line) and the CP-odd $A$ (red dashed line)
Higgs bosons. The values $g_{h\gamma\gamma}=\tilde{g}_{A\gamma\gamma}=0.1$
have been used. The \textit{basic cuts} and the lepton rapidity difference 
cut, cf. Eq. (\ref{cuts3}), have
been applied. \label{distcuts300} }
\end{figure}

In Fig.~\ref{distcuts300},
we present a few kinematic distributions after imposing the 
lepton rapidity difference cut, cf. Eq.~(\ref{cuts3}), together 
with the  \textit{basic cuts}.
First of all, we note that the typical invariant mass of the lepton
pair becomes smaller because the two leptons have a smaller rapidity
separation. Furthermore, the SM Higgs boson rate is suppressed enough
that the CP-odd Higgs boson rate becomes comparable to the SM rate.
From the conservation of total energy, a smaller $m_{e^{-}e^{+}}$
implies a larger energy $E_{\phi}$ (and transverse momentum $p_{T_{\phi}}$)
of the Higgs boson which is evident in Fig.~\ref{distcuts300}. Again,
the $m_{e^{-}e^{+}}$ distribution of 
the CP-even Higgs boson $h$ is similar to the CP-odd
Higgs boson $A$ in the small invariant mass $m_{e^{-}e^{+}}$ region
while similar to the SM Higgs boson $H$ in the large invariant mass
$m_{e^{-}e^{+}}$ region. However,
the detailed distribution depends on the interference pattern generated
by the inclusion of the dimension-6 $h\gamma\gamma$ operator. Clearly,
the difference in the above mentioned kinematical distributions of
the lepton pair and Higgs boson originates from different spin correlations
in the scattering amplitudes for producing the three different Higgs
bosons. To further investigate their difference, we also show in
Fig.~\ref{distcuts300} the distributions of $\cos(\theta_{e^{-}}-\theta_{e^{+}})$
and $\cos(\phi_{e^{-}}-\phi_{e^{+}})$, where $\theta_{i}$ and $\phi_{i}$
are the polar and azimuthal angles of the lepton $i$ in the final
state, and the two new observables $X_{e^{-}e^{+}}$ and $Y_{e^{-}e^{+}}$,
cf. Eq.~(\ref{eq:xydefintion}). The cause of the difference in
the distributions of the above physical observables can be better
understood when we study the case of polarized beams which will be discussed
in the next section. 

Since the polarization of the incoming electron and positron beams
would be possible at the ILC, it is worth analyzing this possibility.
Below we will show that polarized beams are indeed useful for 
distinguishing a CP-even from a CP-odd Higgs boson.

\subsection{Polarized beams}

Let us analyze to what extent the polarization of the collider beams could
help us to discriminate between CP eigenstates of Higgs bosons. 
For simplicity, we assume a $100\%$ degree of 
polarization. Needless to say that in practice the beam polarization
will be only partially longitudinal or partially transverse. Using
the rapidity cuts given in Eq.~(\ref{cuts3}) we calculated the rates
for the $e^{-}_{\lambda_1} e^{+}_{\lambda_2}\to e^{-}e^{+}\phi$ 
(with ${\lambda_1}, {\lambda_2}=L \, {\rm or} \, R$)
processes, which are shown in
Table~\ref{tblpolar} for all the possible polarization states of
the initial leptons. The rates shown in the first row are with the
\textit{basic cuts} only, and the rates in the second row are with
the additional cut on the lepton rapidity
difference, cf. Eq.~(\ref{cuts3}). Since we assume CP is conserved
in this study, the production rate of $e_{R}^{-}e_{R}^{+}$ is the
same as the $e_{L}^{-}e_{L}^{+}$ rate.

\begin{table}

\caption{Cross section for $e^{-}_{\lambda_1} e^{+}_{\lambda_2} \rightarrow e^{-}e^{+}\phi$ when the
initial state leptons are polarized. $L$ and $R$ stand for a left-handed
or right-handed polarized lepton beam. We have taken $m_{\phi}=115\,{\rm GeV}$, 
$g_{h\gamma\gamma}=0.1$
and $\tilde{g}_{A\gamma\gamma}=0.1$. \label{tblpolar}}

\begin{tabular}{c|>{\centering}p{0.4in}|>{\centering}p{0.4in}|>{\centering}p{0.4in}|>{\centering}p{0.4in}|>{\centering}p{0.4in}|>{\centering}p{0.4in}|>{\centering}p{0.4in}|>{\centering}p{0.4in}|>{\centering}p{0.4in}|>{\centering}p{0.4in}|>{\centering}m{0.4in}|>{\centering}p{0.4in}}
\hline 
$\sigma({\textit{fb}})$&
\multicolumn{4}{c|}{$e^{-}_{\lambda_1} e^{+}_{\lambda_2} \rightarrow e^{-}e^{+}H$}&
\multicolumn{4}{c|}{$e^{-}_{\lambda_1} e^{+}_{\lambda_2} \rightarrow e^{-}e^{+}h$}&
\multicolumn{4}{c}{$e^{-}_{\lambda_1} e^{+}_{\lambda_2} \rightarrow e^{-}e^{+}A$}
\tabularnewline
\hline
&
 LL&
 LR&
 RL&
 RR&
 LL&
 LR&
 RL&
 RR&
 LL&
 LR&
 RL&
 RR\tabularnewline
\hline
{\it basic cuts}&
 8.65&
 13.17&
 5.54&
 8.65&
 17.16&
 13.85&
 7.47&
 17.16&
 4.66&
3.77&
 3.77&
 4.66
\tabularnewline
\hline
with $|\eta_{e^{-}}-\eta_{e^{+}}|\leq 3$ &
 1.06&
 1.57&
 0.67&
 1.06& 
 4.93&
 2.33&
 1.85&
 4.93&
 2.21&
 1.61&
 1.61&
 2.21
\tabularnewline
\hline
\end{tabular}
\end{table}

For the sake of illustration we will present below the kinematic distributions
for left-left and left-right handed polarized beams, i.e. $e_{L}^{-}e_{L}^{+}\rightarrow e^{-}e^{+}\phi$
and $e_{L}^{-}e_{R}^{+}\rightarrow e^{-}e^{+}\phi$, denoted as $LL$ and $LR$, respectively.

\subsubsection{Left-left handed polarized beams}

\begin{figure}
\includegraphics[clip,scale=0.7]{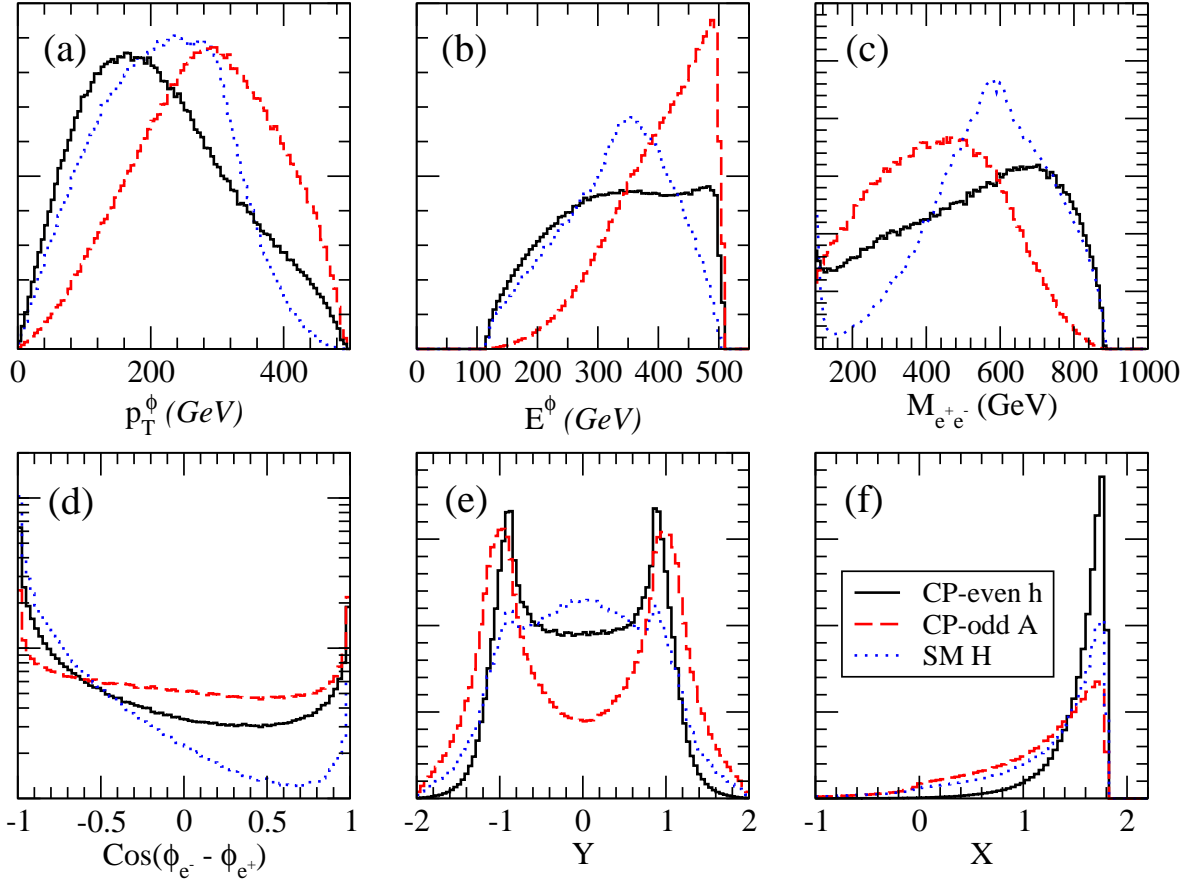}

\caption{Same as Fig.~\ref{distcuts300} for left-left polarized beams.\label{distcuts3mm} }
\end{figure}

\begin{figure}
\includegraphics[scale=0.53]{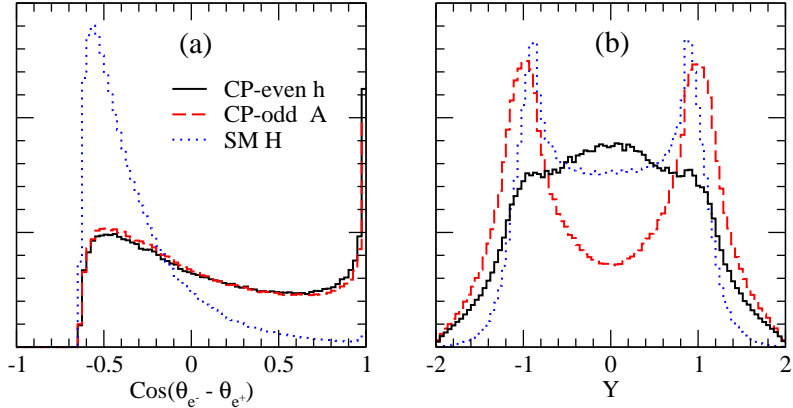}

\caption{
Distributions of $\cos\left(\theta_{e^{-}}-\theta_{e^{+}}\right)$
and $Y_{e^{-}e^{+}}$ after switching off the $hZZ$ interaction in
the $h$ production, namely, only the $h\gamma\gamma $ interaction
is included in this comparison.}

\label{gamma-gamma-h}
\end{figure}

Fig.~\ref{distcuts3mm} shows various kinematic distributions for
$e^{-}_L e^{+}_L \rightarrow e^{-}e^{+}\phi$ when left-left handed polarized
beams are used, in which both the \textit{basic cuts} and the lepton
rapidity difference cut, cf.  Eq.~(\ref{cuts3}), have been
 imposed. It is evident
from the figure that the use of polarized beams enhances the
differences between some distributions, like $\cos\left(\theta_{e^{-}}-\theta_{e^{+}}\right)$
and $Y_{e^{-}e^{+}}$ distributions, cf. Eq.~(\ref{eq:xydefintion}).
We note that for this polarization state of the collider beams there
is no $s$-channel contribution to the production process. The SM
Higgs boson $H$ is produced from $ZZ$ fusion and the CP-odd Higgs
boson $A$ is produced from $\gamma\gamma$ fusion, while the CP-even
Higgs boson $h$ can be produced via both $ZZ$ and $\gamma\gamma$
fusion processes. Through a simple algebra, one can show that \[
Y_{e^{-}e^{+}}\propto\sin\theta_{e^{-}}\sin\theta_{e^{+}}\left[\cos\left(\phi_{e^{-}}+\phi_{e^{+}}\right)-\cos\left(\phi_{e^{-}}-\phi_{e^{-}}\right)\right],\]
therefore, $Y_{e^{-}e^{+}}$ and $\cos\left(\phi_{e^{-}}-\phi_{e^{+}}\right)$
provide almost the same information. To investigate the different
shapes in the $\cos(\phi_{e^{-}}-\phi_{e^{+}})$ and $Y_{e^{-}e^{+}}$
distributions between $h$ and $H$ productions, we show in Fig.~(\ref{gamma-gamma-h})
the distributions of $\cos\left(\theta_{e^{-}}-\theta_{e^{+}}\right)$
and $Y_{e^{-}e^{+}}$ after switching off the $hZZ$ interaction in
the $h$ production, namely, only the $h\gamma\gamma $ interaction
is included in this comparison. It is clear that due to the different
CP property of $h$ and $A$ bosons, the distributions of $Y_{e^{-}e^{+}}$
are very different for $h$ and $A$ bosons, cf. Fig.~\ref{gamma-gamma-h}(b), 
while the distribution of $\cos\left(\theta_{e^{-}}-\theta_{e^{+}}\right)$
is almost the same for $h$ and $A$ bosons, cf. Fig.~\ref{gamma-gamma-h}(a).
Hence, we conclude that the azimuthal angle distributions are more
sensitive to the CP property of the Higgs bosons. Particularly, in
the $Y_{e^{-}e^{+}}$ distribution, a dip occurs around $Y=0$ for
the CP-odd $A$ boson production. In contrast, the rate is enhanced
at $Y=0$ for the CP-even $h$ boson production. Needless to say that
after turning back on the $hZZ$ interaction, the distributions of
the $h$ boson production are also affected by the interference effect
between the $hZZ$ and $h \gamma\gamma $ interactions
which results in the actual distributions shown in Fig.~\ref{distcuts3mm}.
 Similar conclusion
also holds for the $\cos(\phi_{e^{-}}-\phi_{e^{+}})$ distributions. 

\subsubsection{Left-right handed polarized beams}

\begin{figure}
\includegraphics[clip,scale=0.7]{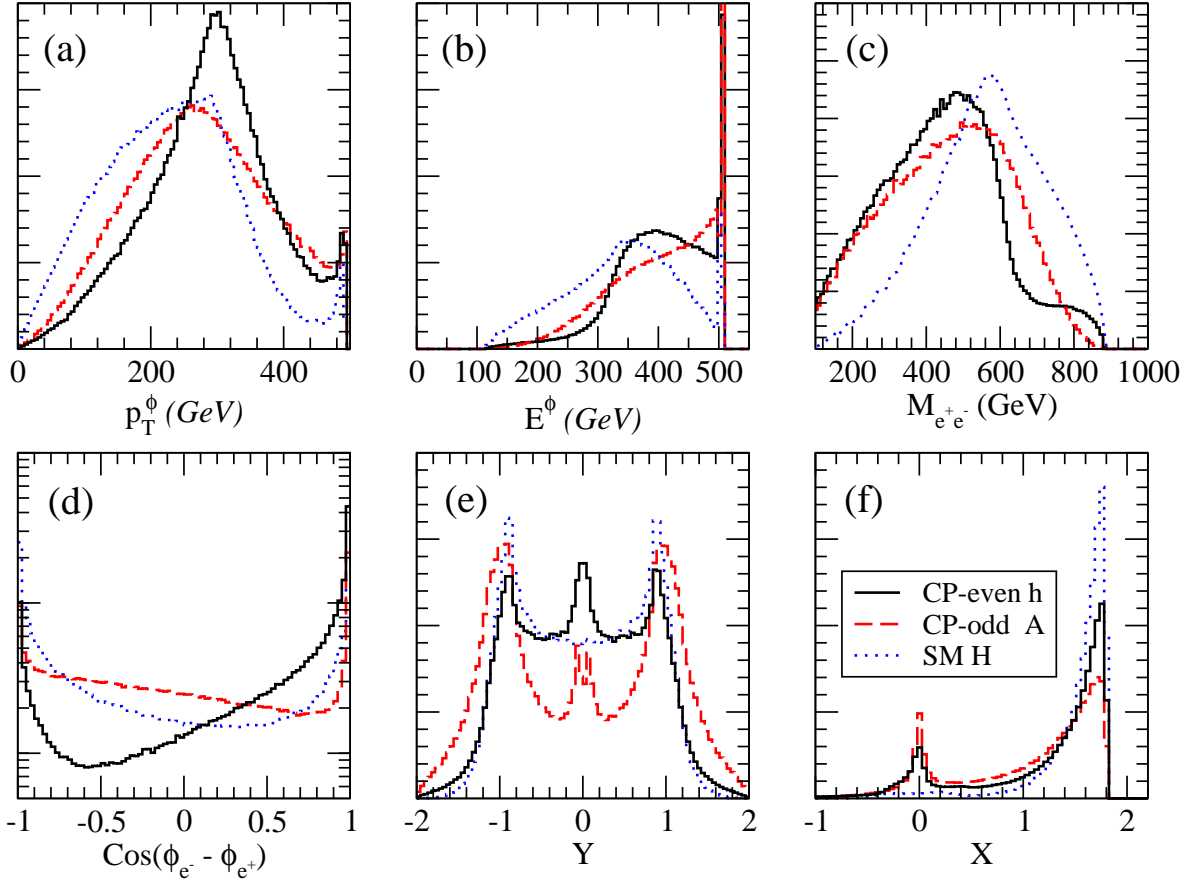}

\caption{Same as Fig.~\ref{distcuts300}, but for left-right polarized beams.
\label{distcuts3mp} }
\end{figure}

Next, we consider the case that the electron beam is left-handedly
polarized while the positron beam is right-handedly polarized. For
this polarization state of the collider beams, both $s$- and $t$-channel
processes contribute to the production of Higgs boson.
In Fig.~\ref{distcuts3mp}, we show the similar kinematic distributions
discussed above for comparison. One distinct feature in this set of
distributions is that there are enhancement rates (spikes) when the
energy $E_{\phi}$ and transverse momentum $p_{T_{\phi}}$ of the
Higgs boson are near 500 GeV, $\cos(\theta_{e^{-}}-\theta_{e^{+}})$
and $\cos(\phi_{e^{-}}-\phi_{e^{+}})$ are near 1, and both $X_{e^{-}e^{+}}$and
$Y_{e^{-}e^{+}}$ are near 0. All of these spikes originate from 
the $s$-channel process. In the $s$-channel production process,
the invariant mass of the di-lepton pair can take a small value, 
$i.e.,$ the di-lepton pair is produced from
the decay of $Z$ boson or a low invariant mass of virtual photon.
Obviously, 
the enhancement factor is larger for a di-lepton pair produced
from a low mass virtual photon than an on-shell $Z$ boson. That explains
why, for example, in the $Y_{e^{-}e^{+}}$ distribution, both $h$
and $A$ production show a strong peak than the $H$ production in
which only the dimension-4 $HZZ$ coupling is considered.

\section{conclusions}

We study the production of Higgs boson ($\phi$) via the $e^{-}e^{+}\rightarrow e^{-}e^{+}\phi$
process at a future International Linear Collider (ILC) with$\sqrt{S}=1\,$TeV.
At this energy the Standard Model 
$ZZ$-fusion rate dominates over the associated
$ZH$ production rate. Here, we
consider a class of theory models based on the effective $\phi VV$
(with $V=Z,\gamma$) couplings generated by dimension-6 operators.
We also show how to use the kinematical
distributions of the final state leptons to discriminate the CP-even
and CP-odd Higgs bosons produced at the ILC, 
when the $\phi\gamma\gamma$ coupling dominates the Higgs boson 
production rate. Assuming the Higgs boson
is identified, we have found that the kinematical
distributions (and correlations) of the final state $e^{-}e^{+}$
pair can be used to test the CP nature of the Higgs boson. In particular,
we have found that requiring the absolute value of their rapidity
difference to be within 3 can greatly reduce the SM rate so that the 
contributions from anomalous couplings are relatively enhanced and it becomes 
easier to use $\phi \gamma\gamma$ couplings to determine the CP property of 
Higgs boson. One
advantage of these cuts is that it is not necessary to determine the
electric charge of the final state $e^{-}e^{+}$
pair. (Particle identification could be challenging for leptons with 
large rapidity values (in magnitude).)
Furthermore, we define two kinematical variables $X_{e^{-}e^{+}}$
and $Y_{e^{-}e^{+}}$. The shapes of these distributions for a CP-even 
Higgs boson are different from those for a CP-odd Higgs boson.
 To distinguish a CP-odd Higgs boson from a CP-even
Higgs boson produced via $e^{-}e^{+}\rightarrow e^{-}e^{+}\phi$ process
at a 1 TeV ILC, it is most effective to use a left-handedly polarized
electron beam colliding a left-handedly polarized positron beam, and
the observable with the biggest discriminating power is the $Y_{e^{-}e^{+}}$
distribution, as defined in Eq.~(\ref{eq:xydefintion}).

\begin{acknowledgments}
We thank T. Tait and A. Belyaev for helpful discussions. The
work of Q.-H.C is supported by the U.S. Department of Energy under
grant No. DE-FG03-94ER40837. G.T.V acknowledges support from SNI and
SEP-PROMEP. F. Larios acknowledges support from Conacyt and the Fulbright-Garcia
Robles fellowship. The work of C.-P. Y. was supported in part by the
U. S. National Science Foundation under award PHY-0244919. 

\newpage
\end{acknowledgments}
\bibliographystyle{apsrev}
\bibliography{reference}

\begin{thebibliography}{33}
\expandafter\ifx\csname natexlab\endcsname\relax\def\natexlab#1{#1}\fi
\expandafter\ifx\csname bibnamefont\endcsname\relax
  \def\bibnamefont#1{#1}\fi
\expandafter\ifx\csname bibfnamefont\endcsname\relax
  \def\bibfnamefont#1{#1}\fi
\expandafter\ifx\csname citenamefont\endcsname\relax
  \def\citenamefont#1{#1}\fi
\expandafter\ifx\csname url\endcsname\relax
  \def\url#1{\texttt{#1}}\fi
\expandafter\ifx\csname urlprefix\endcsname\relax\def\urlprefix{URL }\fi
\providecommand{\bibinfo}[2]{#2}
\providecommand{\eprint}[2][]{\url{#2}}

\bibitem[{\citenamefont{Carena and Haber}(2003)}]{Carena:2002es}
\bibinfo{author}{\bibfnamefont{M.}~\bibnamefont{Carena}} \bibnamefont{and}
  \bibinfo{author}{\bibfnamefont{H.~E.} \bibnamefont{Haber}},
  \bibinfo{journal}{Prog. Part. Nucl. Phys.} \textbf{\bibinfo{volume}{50}},
  \bibinfo{pages}{63} (\bibinfo{year}{2003}), \eprint{hep-ph/0208209}.

\bibitem[{\citenamefont{Gonzalez-Garcia
  et~al.}(1999)\citenamefont{Gonzalez-Garcia, Lietti, and
  Novaes}}]{Gonzalez-Garcia:1998wn}
\bibinfo{author}{\bibfnamefont{M.~C.} \bibnamefont{Gonzalez-Garcia}},
  \bibinfo{author}{\bibfnamefont{S.~M.} \bibnamefont{Lietti}},
  \bibnamefont{and} \bibinfo{author}{\bibfnamefont{S.~F.}
  \bibnamefont{Novaes}}, \bibinfo{journal}{Phys. Rev.}
  \textbf{\bibinfo{volume}{D59}}, \bibinfo{pages}{075008}
  (\bibinfo{year}{1999}), \eprint{hep-ph/9811373}.

\bibitem[{\citenamefont{Kauffman and Desai}(1999)}]{Kauffman:1998yg}
\bibinfo{author}{\bibfnamefont{R.~P.} \bibnamefont{Kauffman}} \bibnamefont{and}
  \bibinfo{author}{\bibfnamefont{S.~V.} \bibnamefont{Desai}},
  \bibinfo{journal}{Phys. Rev.} \textbf{\bibinfo{volume}{D59}},
  \bibinfo{pages}{057504} (\bibinfo{year}{1999}), \eprint{hep-ph/9808286}.

\bibitem[{\citenamefont{Plehn et~al.}(2002)\citenamefont{Plehn, Rainwater, and
  Zeppenfeld}}]{Plehn:2001nj}
\bibinfo{author}{\bibfnamefont{T.}~\bibnamefont{Plehn}},
  \bibinfo{author}{\bibfnamefont{D.~L.} \bibnamefont{Rainwater}},
  \bibnamefont{and}
  \bibinfo{author}{\bibfnamefont{D.}~\bibnamefont{Zeppenfeld}},
  \bibinfo{journal}{Phys. Rev. Lett.} \textbf{\bibinfo{volume}{88}},
  \bibinfo{pages}{051801} (\bibinfo{year}{2002}), \eprint{hep-ph/0105325}.

\bibitem[{\citenamefont{Field}(2002)}]{Field:2002gt}
\bibinfo{author}{\bibfnamefont{B.}~\bibnamefont{Field}},
  \bibinfo{journal}{Phys. Rev.} \textbf{\bibinfo{volume}{D66}},
  \bibinfo{pages}{114007} (\bibinfo{year}{2002}), \eprint{hep-ph/0208262}.

\bibitem[{\citenamefont{Choi et~al.}(2003)\citenamefont{Choi, Miller,
  Muhlleitner, and Zerwas}}]{Choi:2002jk}
\bibinfo{author}{\bibfnamefont{S.~Y.} \bibnamefont{Choi}},
  \bibinfo{author}{\bibfnamefont{D.~J.} \bibnamefont{Miller}},
  \bibinfo{author}{\bibfnamefont{M.~M.} \bibnamefont{Muhlleitner}},
  \bibnamefont{and} \bibinfo{author}{\bibfnamefont{P.~M.}
  \bibnamefont{Zerwas}}, \bibinfo{journal}{Phys. Lett.}
  \textbf{\bibinfo{volume}{B553}}, \bibinfo{pages}{61} (\bibinfo{year}{2003}),
  \eprint{hep-ph/0210077}.

\bibitem[{\citenamefont{Figy and Zeppenfeld}(2004)}]{Figy:2004pt}
\bibinfo{author}{\bibfnamefont{T.}~\bibnamefont{Figy}} \bibnamefont{and}
  \bibinfo{author}{\bibfnamefont{D.}~\bibnamefont{Zeppenfeld}},
  \bibinfo{journal}{Phys. Lett.} \textbf{\bibinfo{volume}{B591}},
  \bibinfo{pages}{297} (\bibinfo{year}{2004}), \eprint{hep-ph/0403297}.

\bibitem[{\citenamefont{Barger et~al.}(1994)\citenamefont{Barger, Cheung,
  Djouadi, Kniehl, and Zerwas}}]{Barger:1993wt}
\bibinfo{author}{\bibfnamefont{V.~D.} \bibnamefont{Barger}},
  \bibinfo{author}{\bibfnamefont{K.-m.} \bibnamefont{Cheung}},
  \bibinfo{author}{\bibfnamefont{A.}~\bibnamefont{Djouadi}},
  \bibinfo{author}{\bibfnamefont{B.~A.} \bibnamefont{Kniehl}},
  \bibnamefont{and} \bibinfo{author}{\bibfnamefont{P.~M.}
  \bibnamefont{Zerwas}}, \bibinfo{journal}{Phys. Rev.}
  \textbf{\bibinfo{volume}{D49}}, \bibinfo{pages}{79} (\bibinfo{year}{1994}),
  \eprint{hep-ph/9306270}.

\bibitem[{\citenamefont{Kilian et~al.}(1996)\citenamefont{Kilian, Kramer, and
  Zerwas}}]{Kilian:1996wu}
\bibinfo{author}{\bibfnamefont{W.}~\bibnamefont{Kilian}},
  \bibinfo{author}{\bibfnamefont{M.}~\bibnamefont{Kramer}}, \bibnamefont{and}
  \bibinfo{author}{\bibfnamefont{P.~M.} \bibnamefont{Zerwas}},
  \bibinfo{journal}{Phys. Lett.} \textbf{\bibinfo{volume}{B381}},
  \bibinfo{pages}{243} (\bibinfo{year}{1996}), \eprint{hep-ph/9603409}.

\bibitem[{\citenamefont{Minkowski}(1998)}]{Minkowski:1997cv}
\bibinfo{author}{\bibfnamefont{P.}~\bibnamefont{Minkowski}},
  \bibinfo{journal}{Int. J. Mod. Phys.} \textbf{\bibinfo{volume}{13}},
  \bibinfo{pages}{2255} (\bibinfo{year}{1998}).

\bibitem[{\citenamefont{Hagiwara et~al.}(2000)\citenamefont{Hagiwara, Ishihara,
  Kamoshita, and Kniehl}}]{Hagiwara:2000tk}
\bibinfo{author}{\bibfnamefont{K.}~\bibnamefont{Hagiwara}},
  \bibinfo{author}{\bibfnamefont{S.}~\bibnamefont{Ishihara}},
  \bibinfo{author}{\bibfnamefont{J.}~\bibnamefont{Kamoshita}},
  \bibnamefont{and} \bibinfo{author}{\bibfnamefont{B.~A.}
  \bibnamefont{Kniehl}}, \bibinfo{journal}{Eur. Phys. J.}
  \textbf{\bibinfo{volume}{C14}}, \bibinfo{pages}{457} (\bibinfo{year}{2000}),
  \eprint{hep-ph/0002043}.

\bibitem[{\citenamefont{Grzadkowski et~al.}(2000)\citenamefont{Grzadkowski,
  Gunion, and Pliszka}}]{Grzadkowski:2000hm}
\bibinfo{author}{\bibfnamefont{B.}~\bibnamefont{Grzadkowski}},
  \bibinfo{author}{\bibfnamefont{J.~F.} \bibnamefont{Gunion}},
  \bibnamefont{and} \bibinfo{author}{\bibfnamefont{J.}~\bibnamefont{Pliszka}},
  \bibinfo{journal}{Nucl. Phys.} \textbf{\bibinfo{volume}{B583}},
  \bibinfo{pages}{49} (\bibinfo{year}{2000}), \eprint{hep-ph/0003091}.

\bibitem[{\citenamefont{Grzadkowski and Pliszka}(2001)}]{Grzadkowski:2000xs}
\bibinfo{author}{\bibfnamefont{B.}~\bibnamefont{Grzadkowski}} \bibnamefont{and}
  \bibinfo{author}{\bibfnamefont{J.}~\bibnamefont{Pliszka}},
  \bibinfo{journal}{Phys. Rev.} \textbf{\bibinfo{volume}{D63}},
  \bibinfo{pages}{115010} (\bibinfo{year}{2001}), \eprint{hep-ph/0012110}.

\bibitem[{\citenamefont{Gunion et~al.}(2003)\citenamefont{Gunion, Haber, and
  Van~Kooten}}]{Gunion:2003fd}
\bibinfo{author}{\bibfnamefont{J.~F.} \bibnamefont{Gunion}},
  \bibinfo{author}{\bibfnamefont{H.~E.} \bibnamefont{Haber}}, \bibnamefont{and}
  \bibinfo{author}{\bibfnamefont{R.}~\bibnamefont{Van~Kooten}},
  \bibinfo{journal}{Linear Collider Physics in the new Milenium, (edited by K.
  Fuji, D. Miller and A. Soni)}  (\bibinfo{year}{2003}),
  \eprint{hep-ph/0301023}.

\bibitem[{\citenamefont{Barger et~al.}(2003)\citenamefont{Barger, Han,
  Langacker, McElrath, and Zerwas}}]{Barger:2003rs}
\bibinfo{author}{\bibfnamefont{V.}~\bibnamefont{Barger}},
  \bibinfo{author}{\bibfnamefont{T.}~\bibnamefont{Han}},
  \bibinfo{author}{\bibfnamefont{P.}~\bibnamefont{Langacker}},
  \bibinfo{author}{\bibfnamefont{B.}~\bibnamefont{McElrath}}, \bibnamefont{and}
  \bibinfo{author}{\bibfnamefont{P.}~\bibnamefont{Zerwas}},
  \bibinfo{journal}{Phys. Rev.} \textbf{\bibinfo{volume}{D67}},
  \bibinfo{pages}{115001} (\bibinfo{year}{2003}), \eprint{hep-ph/0301097}.

\bibitem[{\citenamefont{Desch et~al.}(2004)\citenamefont{Desch, Imhof, Was, and
  Worek}}]{Desch:2003rw}
\bibinfo{author}{\bibfnamefont{K.}~\bibnamefont{Desch}},
  \bibinfo{author}{\bibfnamefont{A.}~\bibnamefont{Imhof}},
  \bibinfo{author}{\bibfnamefont{Z.}~\bibnamefont{Was}}, \bibnamefont{and}
  \bibinfo{author}{\bibfnamefont{M.}~\bibnamefont{Worek}},
  \bibinfo{journal}{Phys. Lett.} \textbf{\bibinfo{volume}{B579}},
  \bibinfo{pages}{157} (\bibinfo{year}{2004}), \eprint{hep-ph/0307331}.

\bibitem[{\citenamefont{Dova and Ferrari}(2005)}]{Dova:2004cj}
\bibinfo{author}{\bibfnamefont{M.~T.} \bibnamefont{Dova}} \bibnamefont{and}
  \bibinfo{author}{\bibfnamefont{S.}~\bibnamefont{Ferrari}},
  \bibinfo{journal}{Phys. Lett.} \textbf{\bibinfo{volume}{B605}},
  \bibinfo{pages}{376} (\bibinfo{year}{2005}), \eprint{hep-ph/0406313}.

\bibitem[{\citenamefont{Rouge}(2005)}]{Rouge:2005iy}
\bibinfo{author}{\bibfnamefont{A.}~\bibnamefont{Rouge}},
  \bibinfo{journal}{Phys. Lett.} \textbf{\bibinfo{volume}{B619}},
  \bibinfo{pages}{43} (\bibinfo{year}{2005}), \eprint{hep-ex/0505014}.

\bibitem[{\citenamefont{Ellis et~al.}(2005{\natexlab{a}})\citenamefont{Ellis,
  Lee, and Pilaftsis}}]{Ellis:2005ik}
\bibinfo{author}{\bibfnamefont{J.~R.} \bibnamefont{Ellis}},
  \bibinfo{author}{\bibfnamefont{J.~S.} \bibnamefont{Lee}}, \bibnamefont{and}
  \bibinfo{author}{\bibfnamefont{A.}~\bibnamefont{Pilaftsis}},
  \bibinfo{journal}{Phys. Rev.} \textbf{\bibinfo{volume}{D72}},
  \bibinfo{pages}{095006} (\bibinfo{year}{2005}{\natexlab{a}}),
  \eprint{hep-ph/0507046}.

\bibitem[{\citenamefont{Biswal et~al.}(2006)\citenamefont{Biswal, Godbole,
  Singh, and Choudhury}}]{Biswal:2005fh}
\bibinfo{author}{\bibfnamefont{S.~S.} \bibnamefont{Biswal}},
  \bibinfo{author}{\bibfnamefont{R.~M.} \bibnamefont{Godbole}},
  \bibinfo{author}{\bibfnamefont{R.~K.} \bibnamefont{Singh}}, \bibnamefont{and}
  \bibinfo{author}{\bibfnamefont{D.}~\bibnamefont{Choudhury}},
  \bibinfo{journal}{Phys. Rev.} \textbf{\bibinfo{volume}{D73}},
  \bibinfo{pages}{035001} (\bibinfo{year}{2006}), \eprint{hep-ph/0509070}.

\bibitem[{\citenamefont{Han and Jiang}(2001)}]{Han:2000mi}
\bibinfo{author}{\bibfnamefont{T.}~\bibnamefont{Han}} \bibnamefont{and}
  \bibinfo{author}{\bibfnamefont{J.}~\bibnamefont{Jiang}},
  \bibinfo{journal}{Phys. Rev.} \textbf{\bibinfo{volume}{D63}},
  \bibinfo{pages}{096007} (\bibinfo{year}{2001}), \eprint{hep-ph/0011271}.

\bibitem[{\citenamefont{Grzadkowski and Gunion}(1992)}]{Grzadkowski:1992sa}
\bibinfo{author}{\bibfnamefont{B.}~\bibnamefont{Grzadkowski}} \bibnamefont{and}
  \bibinfo{author}{\bibfnamefont{J.~F.} \bibnamefont{Gunion}},
  \bibinfo{journal}{Phys. Lett.} \textbf{\bibinfo{volume}{B294}},
  \bibinfo{pages}{361} (\bibinfo{year}{1992}), \eprint{hep-ph/9206262}.

\bibitem[{\citenamefont{Gounaris and Tsirigoti}(1997)}]{Gounaris:1997ef}
\bibinfo{author}{\bibfnamefont{G.~J.} \bibnamefont{Gounaris}} \bibnamefont{and}
  \bibinfo{author}{\bibfnamefont{G.~P.} \bibnamefont{Tsirigoti}},
  \bibinfo{journal}{Phys. Rev.} \textbf{\bibinfo{volume}{D56}},
  \bibinfo{pages}{3030} (\bibinfo{year}{1997}), \eprint{hep-ph/9703446}.

\bibitem[{\citenamefont{Ellis et~al.}(2005{\natexlab{b}})\citenamefont{Ellis,
  Lee, and Pilaftsis}}]{Ellis:2004hw}
\bibinfo{author}{\bibfnamefont{J.~R.} \bibnamefont{Ellis}},
  \bibinfo{author}{\bibfnamefont{J.~S.} \bibnamefont{Lee}}, \bibnamefont{and}
  \bibinfo{author}{\bibfnamefont{A.}~\bibnamefont{Pilaftsis}},
  \bibinfo{journal}{Nucl. Phys.} \textbf{\bibinfo{volume}{B718}},
  \bibinfo{pages}{247} (\bibinfo{year}{2005}{\natexlab{b}}),
  \eprint{hep-ph/0411379}.

\bibitem[{\citenamefont{Niezurawski et~al.}(2005)\citenamefont{Niezurawski,
  Zarnecki, and Krawczyk}}]{Niezurawski:2004ga}
\bibinfo{author}{\bibfnamefont{P.}~\bibnamefont{Niezurawski}},
  \bibinfo{author}{\bibfnamefont{A.~F.} \bibnamefont{Zarnecki}},
  \bibnamefont{and} \bibinfo{author}{\bibfnamefont{M.}~\bibnamefont{Krawczyk}},
  \bibinfo{journal}{Acta Phys. Polon.} \textbf{\bibinfo{volume}{B36}},
  \bibinfo{pages}{833} (\bibinfo{year}{2005}), \eprint{hep-ph/0410291}.

\bibitem[{\citenamefont{Gunion et~al.}(1998)\citenamefont{Gunion, Han, and
  Sobey}}]{Gunion:1998jc}
\bibinfo{author}{\bibfnamefont{J.~F.} \bibnamefont{Gunion}},
  \bibinfo{author}{\bibfnamefont{T.}~\bibnamefont{Han}}, \bibnamefont{and}
  \bibinfo{author}{\bibfnamefont{R.}~\bibnamefont{Sobey}},
  \bibinfo{journal}{Phys. Lett.} \textbf{\bibinfo{volume}{B429}},
  \bibinfo{pages}{79} (\bibinfo{year}{1998}), \eprint{hep-ph/9801317}.

\bibitem[{\citenamefont{Weinberg}(1979)}]{Weinberg:1978kz}
\bibinfo{author}{\bibfnamefont{S.}~\bibnamefont{Weinberg}},
  \bibinfo{journal}{Physica} \textbf{\bibinfo{volume}{A96}},
  \bibinfo{pages}{327} (\bibinfo{year}{1979}).

\bibitem[{\citenamefont{Georgi}(1991)}]{Georgi:1991ch}
\bibinfo{author}{\bibfnamefont{H.}~\bibnamefont{Georgi}},
  \bibinfo{journal}{Nucl. Phys.} \textbf{\bibinfo{volume}{B361}},
  \bibinfo{pages}{339} (\bibinfo{year}{1991}).

\bibitem[{\citenamefont{Buchmuller and Wyler}(1986)}]{Buchmuller:1985jz}
\bibinfo{author}{\bibfnamefont{W.}~\bibnamefont{Buchmuller}} \bibnamefont{and}
  \bibinfo{author}{\bibfnamefont{D.}~\bibnamefont{Wyler}},
  \bibinfo{journal}{Nucl. Phys.} \textbf{\bibinfo{volume}{B268}},
  \bibinfo{pages}{621} (\bibinfo{year}{1986}).

\bibitem[{\citenamefont{Arzt et~al.}(1995)\citenamefont{Arzt, Einhorn, and
  Wudka}}]{Arzt:1994gp}
\bibinfo{author}{\bibfnamefont{C.}~\bibnamefont{Arzt}},
  \bibinfo{author}{\bibfnamefont{M.~B.} \bibnamefont{Einhorn}},
  \bibnamefont{and} \bibinfo{author}{\bibfnamefont{J.}~\bibnamefont{Wudka}},
  \bibinfo{journal}{Nucl. Phys.} \textbf{\bibinfo{volume}{B433}},
  \bibinfo{pages}{41} (\bibinfo{year}{1995}), \eprint{hep-ph/9405214}.

\bibitem[{\citenamefont{Hagiwara et~al.}(1993)\citenamefont{Hagiwara,
  Szalapski, and Zeppenfeld}}]{Hagiwara:1993qt}
\bibinfo{author}{\bibfnamefont{K.}~\bibnamefont{Hagiwara}},
  \bibinfo{author}{\bibfnamefont{R.}~\bibnamefont{Szalapski}},
  \bibnamefont{and}
  \bibinfo{author}{\bibfnamefont{D.}~\bibnamefont{Zeppenfeld}},
  \bibinfo{journal}{Phys. Lett.} \textbf{\bibinfo{volume}{B318}},
  \bibinfo{pages}{155} (\bibinfo{year}{1993}), \eprint{hep-ph/9308347}.

\bibitem[{\citenamefont{Hagiwara and Zeppenfeld}(1986)}]{Hagiwara:1985yu}
\bibinfo{author}{\bibfnamefont{K.}~\bibnamefont{Hagiwara}} \bibnamefont{and}
  \bibinfo{author}{\bibfnamefont{D.}~\bibnamefont{Zeppenfeld}},
  \bibinfo{journal}{Nucl. Phys.} \textbf{\bibinfo{volume}{B274}},
  \bibinfo{pages}{1} (\bibinfo{year}{1986}).

\bibitem[{\citenamefont{Carlson}(PhD thesis)}]{Carlson:1995ei}
\bibinfo{author}{\bibfnamefont{D.~O.} \bibnamefont{Carlson}},
  \bibinfo{journal}{Physics of single-top quark production at hadron colliders}
   (\bibinfo{year}{PhD thesis}), \eprint{hep-ph/9508278}.

\end{thebibliography}
\newpage

\appendix

\section{Helicity Amplitudes for $e^{-}e^{+}\rightarrow e^{-}e^{+}\phi$ \label{sec:Helicity-Amplitudes}}

In this appendix we outline the calculation of the helicity amplitudes
for $e^{-}e^{+}\rightarrow e^{-}e^{+}\phi$. We use the method of
Ref.~\cite{Hagiwara:1985yu,Carlson:1995ei}, which breaks down the four-dimensional
Dirac algebra into an equivalent two-dimensional one. In the Weyl
basis, Dirac spinors have the form \begin{equation}
\left(\begin{array}{c}
\psi_{+}\\
\psi_{-}\end{array}\right),\end{equation}
 where \begin{equation}
\psi_{\pm}=\left\{ \begin{array}{cc}
u_{\pm}^{(\lambda=1)}=\omega_{\pm}\chi_{1/2}\\
u_{\pm}^{(\lambda=-)}=\omega_{\mp}\chi_{-1/2}\end{array}\right.\end{equation}
 for fermions and \begin{equation}
\psi_{\pm}=\left\{ \begin{array}{cc}
v_{\pm}^{(\lambda=1)}=\pm\omega_{\mp}\chi_{-1/2}\\
v_{\pm}^{(\lambda=-)}=\mp\omega_{\pm}\chi_{1/2}\end{array}\right.\end{equation}
 for antifermions. $\omega_{\pm}=\sqrt{E\pm\left|\vec{p}\right|}$
and $\chi_{\lambda/2}$ stands for the eigenvectors of the helicity
operator $\hat{p}\cdot\vec{\sigma}$ with eigenvalue $\lambda=1$
for spin-up fermions and $\lambda=-1$ for spin-down fermions: \begin{equation}
\chi_{1/2}=\left(\begin{array}{c}
\cos\theta/2\\
e^{i\phi}\sin\theta/2\end{array}\right),\,\,\chi_{-1/2}=\left(\begin{array}{c}
-e^{i\phi}\sin\theta/2\\
\cos\theta/2\end{array}\right).\end{equation}
 We will use the shorthand notation $|p_{i}\pm\!\!\!>$ for $\chi_{\pm1/2}$.
In this basis, the Dirac matrices have the form \begin{equation}
\gamma^{0}=\left(\begin{array}{cc}
0 & 1\\
1 & 0\end{array}\right),\,\,\,\gamma^{j}=\left(\begin{array}{cc}
0 & -\sigma_{j}\\
\sigma_{j} & 0\end{array}\right),\,\,\,\gamma^{5}=\left(\begin{array}{cc}
1 & 0\\
0 & -1\end{array}\right),\end{equation}
 where $\sigma_{j}$ are the $2\times2$ Pauli matrices. We can write
\begin{equation}
\not\! p=\left(\begin{array}{cc}
0 & p_{0}+\vec{\sigma}\cdot\vec{p}\\
p_{0}-\vec{\sigma}\cdot\vec{p} & 0\end{array}\right)\equiv\left(\begin{array}{cc}
0 & \not\! p_{+}\\
\not\! p_{-} & 0\end{array}\right),\end{equation}
 where $\gamma_{\pm}^{\mu}=(1,\mp\vec{\sigma})$. Let us denote the
helicity amplitudes by 
$\mathcal{M}(\lambda_{e_{1}^{-}},\lambda_{e_{2}^{+}},\lambda_{e_{3}^{-}},\lambda_{e_{4}^{+}})$,
where $\lambda_{i}$ represents the helicity of the particle $i$.
The convention for the momenta of the external particles is given
in Eq.~(\ref{momenta}). After a straightforward calculation, the following
helicity amplitudes 
for $e^{-}e^{+}\rightarrow e^{-}e^{+}\phi$ are
obtained.

\subsection{SM Higgs boson\label{sub:SM-Higgs-boson}}

\begin{figure}
\includegraphics[scale=0.4]{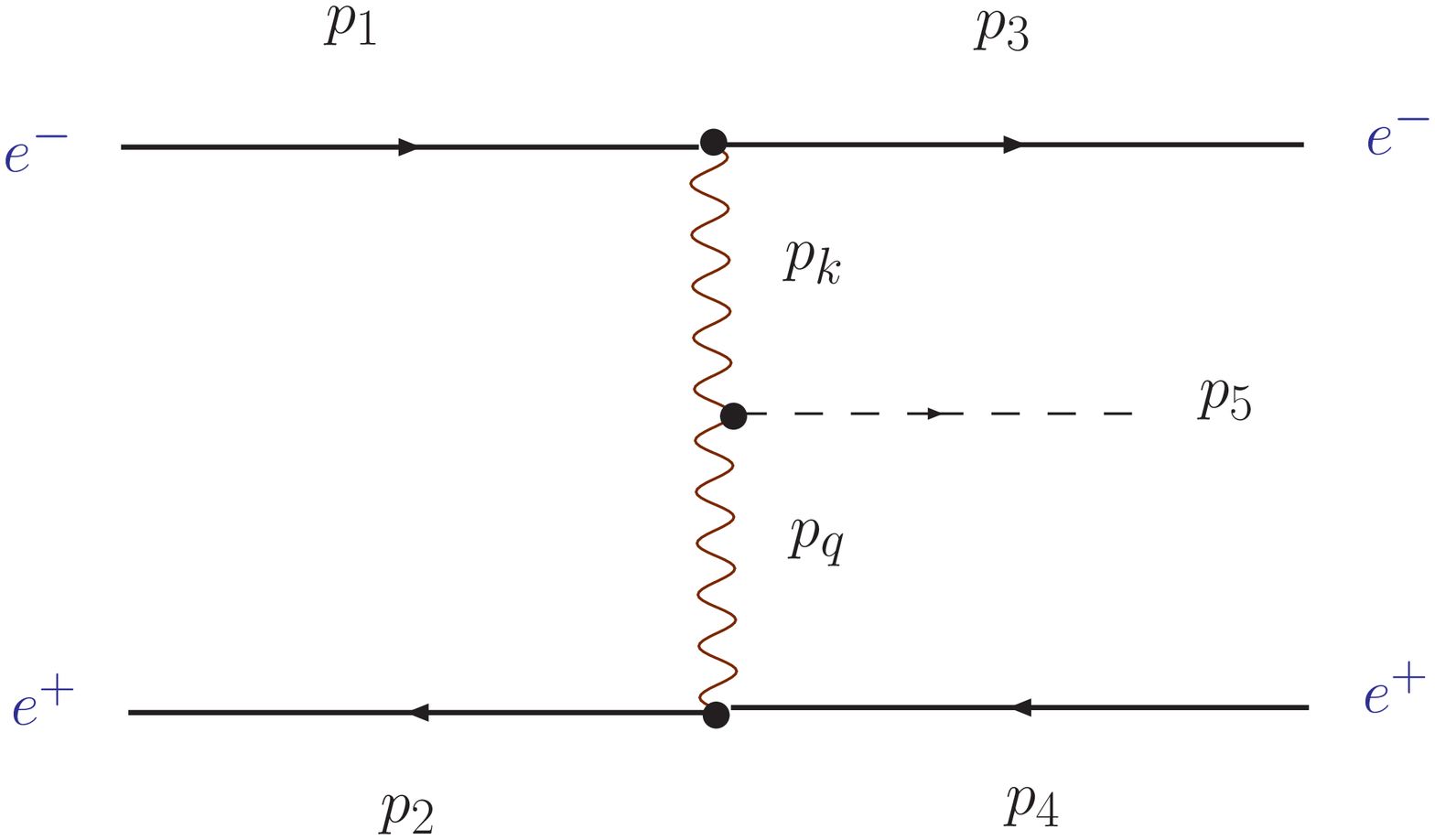}~~~~\includegraphics[scale=0.4]{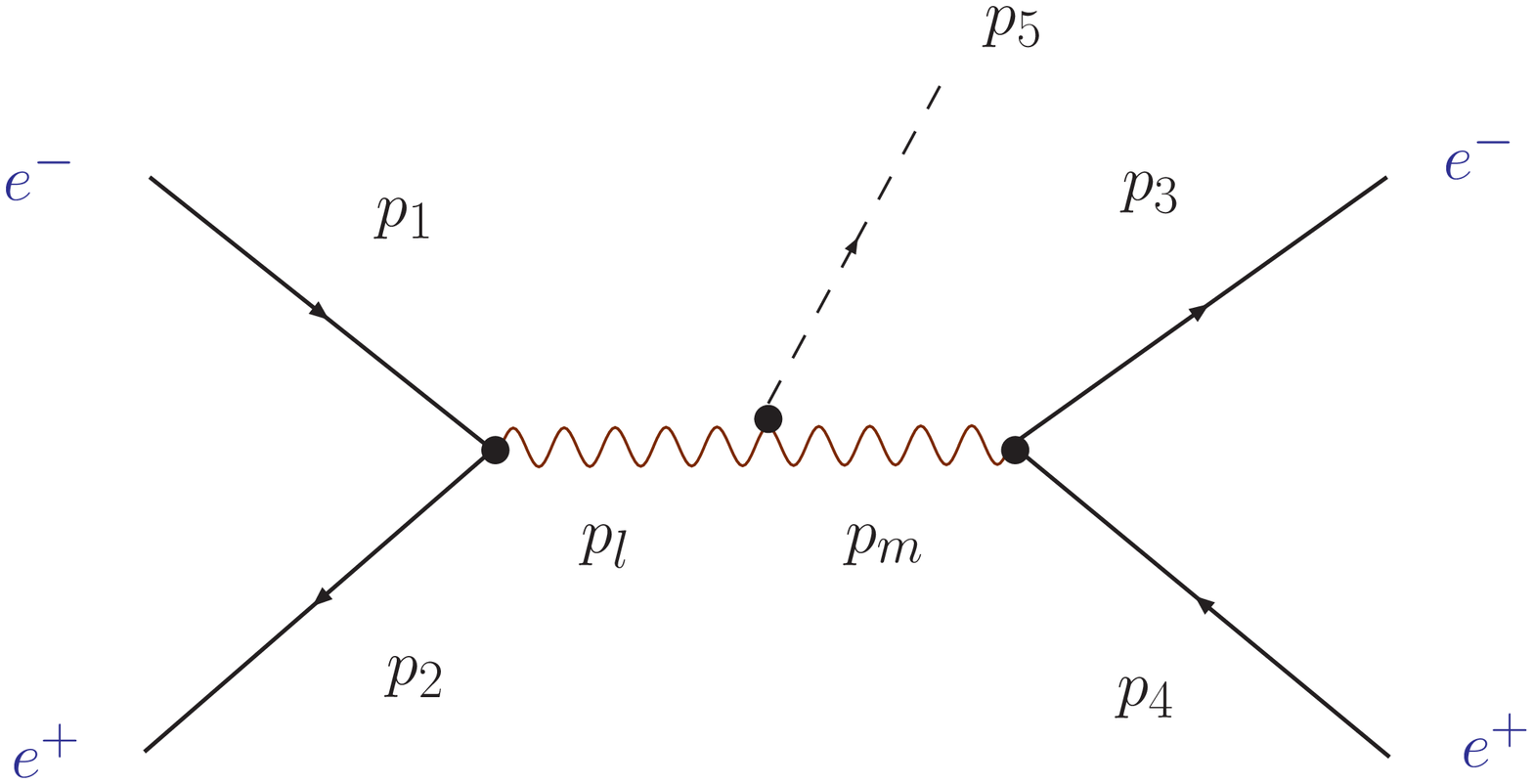}

\caption{Feynman diagrams for $e^{-}e^{+}\rightarrow e^{-}e^{+}\phi$.\label{feynman:channels}}
\end{figure}

The SM Higgs boson can be produced through the Higgstrahlung or $ZZ$-fusion
process, as shown in Fig.~\ref{feynman:channels}. The former is
$s$-channel like but the latter is $t$-channel like. The helicity
amplitudes 
$\mathcal{M}_{SM}^{HZZ}
(\lambda_{e_{1}^{-}},\lambda_{e_{2}^{+}},\lambda_{e_{3}^{-}},\lambda_{e_{4}^{+}})$
for the SM Higgs boson production via 
$
e^{-}(p_1,\lambda_{e_{1}^{-}}) 
e^{+}(p_2,\lambda_{e_{2}^{+}}) 
\to
e^{-}(p_3,\lambda_{e_{3}^{-}}) 
e^{+}(p_4,\lambda_{e_{4}^{+}}) 
H(p_5)
$
are listed as follows:

\begin{eqnarray}
\mathcal{M}_{SM}^{HZZ}(++++) & = & C_{fac}\frac{2\, g_{HZZ}e_{L}^{Z}e_{R}^{Z}}{(p_{k}^{2}-m_{Z}^{2})(p_{q}^{2}-m_{Z}^{2})}
\BKM{1}{2}\BKP{4}{3} \, , \end{eqnarray}
\begin{eqnarray}
\mathcal{M}_{SM}^{HZZ}(----) & = & C_{fac}\frac{2\, g_{HZZ}e_{L}^{Z}e_{R}^{Z}}{(p_{k}^{2}-m_{Z}^{2})(p_{q}^{2}-m_{Z}^{2})}
\BKM{1}{4}\BKP{2}{3} \, , \end{eqnarray}
\begin{eqnarray}
\mathcal{M}_{SM}^{HZZ}(-++-) & = & C_{fac}\frac{2\, g_{HZZ}e_{L}^{Z}e_{R}^{Z}}{(p_{l}^{2}-m_{Z}^{2})(p_{m}^{2}-m_{Z}^{2})}
\BKM{2}{4}\BKP{3}{1} \, , \end{eqnarray}
\begin{eqnarray}
\mathcal{M}_{SM}^{HZZ}(+--+) & = & C_{fac}\frac{2\, g_{HZZ}e_{L}^{Z}e_{R}^{Z}}{(p_{l}^{2}-m_{Z}^{2})(p_{m}^{2}-m_{Z}^{2})}
\BKP{2}{4}\BKM{3}{1} \, , \end{eqnarray}
\begin{eqnarray}
\mathcal{M}_{SM}^{HZZ}(-+-+) & = & C_{fac}\Biggl\{\frac{2\, g_{HZZ}e_{L}^{Z}e_{R}^{Z}}{(p_{k}^{2}-m_{Z}^{2})(p_{q}^{2}-m_{Z}^{2})}\BKP{4}{3}\BKM{1}{2}\nonumber \\
 &  & \,\,\,\,\,\,\,\,\,\,\,\,\,\,+\frac{2\, g_{HZZ}e_{R}^{Z}e_{L}^{Z}}{(p_{l}^{2}-m_{Z}^{2})(p_{m}^{2}-m_{Z}^{2})}\BKM{2}{3}
\BKP{4}{1}\Biggr{\}} \, , \end{eqnarray}
\begin{eqnarray}
\mathcal{M}_{SM}^{HZZ}(+-+-) & = & C_{fac}\Biggl\{\frac{2\, g_{HZZ}e_{L}^{Z}e_{R}^{Z}}{(p_{k}^{2}-m_{Z}^{2})(p_{q}^{2}-m_{Z}^{2})}\BKP{1}{2}\BKM{4}{3}\nonumber \\
 &  & \,\,\,\,\,\,\,\,\,\,\,\,\,\,+\frac{2\, g_{HZZ}e_{L}^{Z}e_{R}^{Z}}{(p_{l}^{2}-m_{Z}^{2})(p_{m}^{2}-m_{Z}^{2})}
\BKM{2}{3}\BKP{4}{1}\Biggr{\}} \, . \end{eqnarray}
 Here, \begin{eqnarray*}
e_{L}^{Z} & = & {\displaystyle \frac{1}{c_{W}}(-\frac{1}{2}+s_{W}^{2})},\\
e_{R}^{Z} & = & -s_{W},\end{eqnarray*}
 and the SM $H$-$Z$-$Z$ coupling is\[
g_{HZZ}=\frac{g}{c_{W}}\, m_{Z},\]
in which $s_W=\sin \theta_W$ and $c_W=\cos \theta_W$ with $\theta_W$ being 
the weak mixing angle. 
The common factor $C_{fac}$ is defined as \[
C_{fac}=(-ig^{2})\sqrt{2E_{1}}\sqrt{2E_{2}}\sqrt{2E_{3}}\sqrt{2E_{4}} \, , \]
and the momenta of the intermediate state
particles are defined as $p_{q}=p_{2}-p_{4}$, $p_{k}=p_{1}-p_{3}$, $p_{l}=p_{1}+p_{2}$,
and $p_{m}=p_{3}+p_{4}$. 

\subsection{CP-even Higgs Boson}

The CP-even Higgs boson gets both the SM-like contribution and the
anomalous contribution,\[
\mathcal{M}_{even}=\mathcal{M}_{SM}^{hZZ}+\sum_{V=\gamma,Z}\mathcal{M}^{hVV}.\]
The SM-like contribution can easily obtained from Sec.~\ref{sub:SM-Higgs-boson}
by replacing the SM Higgs boson ($H$) with the CP-even Higgs boson
($h$). The anomalous contributions are listed as follows:

\begin{flushleft}\begin{eqnarray}
\mathcal{M}^{hVV}(++++) & = & C_{fac}\frac{g_{hVV}e_{L}^{V}e_{R}^{V}}{p_{k}^{2}p_{q}^{2}}\Biggl\{2(p_{k}\cdot p_{q})\BKP{1}{4}\BKM{2}{3}-\BSKP{1}{p_{q}}{3}\BSKM{2}{p_{k}}{4}\Biggr{\}},\nonumber \\
\\\mathcal{M}^{hVV}(----) & = & C_{fac}\frac{g_{hVV}e_{L}^{V}e_{R}^{V}}{p_{k}^{2}p_{q}^{2}}\Biggl\{2(p_{k}\cdot p_{q})\BKM{1}{4}\BKP{2}{3}-\BSKM{2}{p_{q}}{1}\BSKP{3}{p_{k}}{4}\Biggr{\}},\nonumber \\
\\\mathcal{M}^{hVV}(-++-) & = & C_{fac}\frac{g_{hVV}e_{L}^{\gamma V}e_{R}^{V}}{p_{l}^{2}p_{m}^{2}}\Biggl\{2(p_{l}\cdot p_{m})\BKM{2}{4}\BKP{3}{1}-\BSKM{2}{p_{m}}{1}\BSKP{3}{p_{l}}{4}\Biggr{\}},\nonumber \\
\\\mathcal{M}^{hVV}(+--+) & = & C_{fac}\frac{g_{hVV}e_{L}^{V}e_{R}^{V}}{p_{l}^{2}p_{m}^{2}}\Biggl\{2(p_{l}\cdot p_{m})\BKP{2}{4}\BKM{3}{1}-\BSKP{2}{p_{m}}{1}\BSKM{3}{p_{l}}{4}\Biggr{\}},\nonumber \\
\\\mathcal{M}^{hVV}(-+-+) & = & C_{fac}\frac{g_{hVV}e_{L}^{V}e_{R}^{V}}{p_{k}^{2}p_{q}^{2}}\Biggl\{2(p_{k}\cdot p_{q})\BKP{4}{3}\BKM{1}{2}-\BSKM{2}{p_{q}}{1}\BSKM{3}{p_{k}}{4}\Biggr{\}}\nonumber \\
\\ & + & C_{fac}\frac{g_{hVV}e_{L}^{V}e_{R}^{V}}{p_{l}^{2}p_{m}^{2}}\Biggl\{2(p_{l}\cdot p_{m})\BKM{2}{3}\BKP{4}{1}-\BSKM{2}{p_{m}}{1}\BSKM{3}{p_{l}}{4}\Biggr{\}},\nonumber \\
\\\mathcal{M}^{hVV}(+-+-) & = & C_{fac}\frac{g_{hVV}e_{L}^{V}e_{R}^{V}}{p_{k}^{2}p_{q}^{2}}\Biggl\{2(p_{k}\cdot p_{q})\BKP{1}{2}\BKM{4}{3}-\BSKP{1}{p_{q}}{3}\BSKP{2}{p_{k}}{4}\Biggr{\}}\nonumber \\
 & + & C_{fac}\frac{g_{hVV}e_{L}^{V}e_{R}^{V}}{p_{l}^{2}p_{m}^{2}}\Biggl\{2(p_{l}\cdot p_{m})
\BKP{2}{3}\BKM{4}{1}-\BSKP{2}{p_{m}}{1}\BSKP{3}{p_{l}}{4}\Biggr{\}} \, . \nonumber \end{eqnarray}
 Here,\begin{eqnarray*}
e_{L}^{V} & = & \left\{ \begin{array}{ccc}
{\displaystyle \frac{1}{c_{W}}(-\frac{1}{2}+s_{W}^{2})} &  & V=Z\\
-s_{W} &  & V=\gamma\end{array}\right.,\\
e_{R}^{V} & = & \left\{ \begin{array}{ccc}
{\displaystyle \frac{s_{W}^{2}}{c_{W}}} & \,\,\,\,\,\,\,\,\,\,\,\,\,\,\,\,\,\,\,\,\,\,\,\,\,\, & V=Z\\
-s_{W} &  & V=\gamma\end{array}\right.,\end{eqnarray*}
 and the anomalous
couplings $g_{hVV}$ are related to the coefficients $f_{i}$
in Eq.~(\ref{eq: LagphiAA}) by\[
g_{hVV}=-g\frac{s_{W}^{2}m_{W}}{2\Lambda^{2}}\left(f_{BB}+f_{WW}\right).\]
Again, we have defined $p_{q}=p_{2}-p_{4}$, $p_{k}=p_{1}-p_{3}$, $p_{l}=p_{1}+p_{2}$,
and $p_{m}=p_{4}+p_{3}$. \par\end{flushleft}

\subsection{CP-odd Higgs Boson}

The CP-odd Higgs boson receives contributions only from the anomalous
couplings, i.e.\[
\mathcal{M}_{odd}=\sum_{V=\gamma,Z}\mathcal{M}^{AVV}.\]
Using the method of Ref. \cite{Kauffman:1998yg}, we 
obtained the following helicity amplitudes: \begin{eqnarray*}
\mathcal{M}^{AVV}(++++) & = & C_{fac}\frac{{\widetilde{g}}_{AVV}e_{L}^{V}e_{R}^{V}}{p_{k}^{2}p_{q}^{2}}\Biggl\{\BKP{1}{4}\BSSKM{2}{p_{k}}{p_{q}}{3}-\BKM{3}{2}\BSSKP{4}{p_{k}}{p_{q}}{1}\Biggr\},\end{eqnarray*}
\begin{eqnarray*}
\mathcal{M}^{AVV}(----) & = & C_{fac}\frac{{\widetilde{g}}_{AVV}e_{L}^{V}e_{R}^{V}}{p_{k}^{2}p_{q}^{2}}\Biggl\{\BKP{3}{2}\BSSKM{4}{p_{k}}{p_{q}}{1}-\BKM{1}{4}\BSSKP{2}{p_{k}}{p_{q}}{3}\Biggr\},\end{eqnarray*}
\begin{eqnarray*}
\mathcal{M}^{AVV}(-++-) & = & C_{fac}\frac{{\widetilde{g}}_{AVV}e_{L}^{V}e_{R}^{V}}{p_{l}^{2}p_{m}^{2}}\Biggl\{\BKP{1}{3}\BSSKM{4}{p_{l}}{p_{m}}{2}-\BKM{2}{4}\BSSKP{3}{p_{l}}{p_{m}}{1}\Biggr\},\end{eqnarray*}
\begin{eqnarray*}
\mathcal{M}^{AVV}(+--+) & = & C_{fac}\frac{{\widetilde{g}}_{AVV}e_{L}^{V}e_{R}^{V}}{p_{l}^{2}p_{m}^{2}}\Biggl\{\BKP{2}{4}\BSSKM{3}{p_{l}}{p_{m}}{1}-\BKM{1}{3}\BSSKP{4}{p_{l}}{p_{m}}{2}\Biggr\},\end{eqnarray*}
\begin{eqnarray*}
\mathcal{M}^{AVV}(-+-+) & = & C_{fac}\frac{{\widetilde{g}}_{AVV}e_{L}^{V}e_{R}^{V}}{p_{k}^{2}p_{q}^{2}}\Biggl\{\BKP{3}{4}\BSSKM{2}{p_{k}}{p_{q}}{1}-\BKM{1}{2}\BSSKP{4}{p_{k}}{p_{q}}{3}\Biggr\}\\
 & + & C_{fac}\frac{{\widetilde{g}}_{AVV}e_{L}^{V}e_{R}^{V}}{p_{l}^{2}p_{m}^{2}}\Biggl\{\BKP{1}{4}\BSSKM{3}{p_{l}}{p_{m}}{2}-\BKM{2}{3}\BSSKP{4}{p_{l}}{p_{m}}{1}\Biggr\},\end{eqnarray*}
\begin{eqnarray*}
\mathcal{M}^{AVV}(+-+-) & = & C_{fac}\frac{{\widetilde{g}}_{AVV}e_{L}^{V}e_{R}^{V}}{p_{k}^{2}p_{q}^{2}}\Biggl\{\BKP{1}{2}\BSSKM{4}{p_{k}}{p_{q}}{3}-\BKM{3}{4}\BSSKP{2}{p_{k}}{p_{q}}{1}\Biggr\}\\
 & + & C_{fac}\frac{{\widetilde{g}}_{AVV}e_{L}^{V}e_{R}^{V}}{p_{l}^{2}p_{m}^{2}}\Biggl\{\BKP{2}{3}\BSSKM{4}{p_{l}}{p_{m}}{1}-\BKM{1}{4}\BSSKP{3}{p_{l}}{p_{m}}{2}\Biggr\},\end{eqnarray*}
where the anomalous couplings $\tilde{g}_{AVV}$ are related to 
the coefficients $f_{i}$ in Eq.~(\ref{eq: LagphiAA}) by\[
\tilde{g}_{AVV}=-g\frac{s_{W}^{2}m_{W}}{2\Lambda^{2}}\left(\tilde{f}_{BB}+\tilde{f}_{WW}\right).\]

\end{document}